\documentclass[preprint,authoryear,12pt]{elsarticle}  % for submission!

\usepackage{graphicx}
\usepackage{amsmath}
\usepackage{epsfig}
\usepackage[latin1]{inputenc}
\usepackage{natbib}
\usepackage{amssymb}
\usepackage{subcaption}
\usepackage[export]{adjustbox}
\usepackage[T1]{fontenc}
\usepackage{ae,aecompl,wasysym}
\usepackage{multicol}
\usepackage{url}
\usepackage{hyperref}

\title{Boulder Stranding in Ejecta Launched by an Impact Generated Seismic Pulse}
\author[a1]{Esteban Wright\corref{cor1}}
	\ead{ewrig15@ur.rochester.edu}
\author[a1]{Alice C. Quillen} 
	\ead{alice.quillen@rochester.edu}
\author[a1]{Juliana South}
	\ead{jsouth@u.rochester.edu}
\author[a2]{Randal C. Nelson}  
	\ead{nelson@cs.rochester.edu}
\author[a3]{Paul S\'anchez}
	\ead{diego.sanchez-lana@colorado.edu}
\author[a1,a4]{Larkin Martini}
	\ead{lmartini@mymail.mines.edu}
\author[a5,a6]{Stephen R. Schwartz}
	\ead{srs51@email.arizona.edu}
\author[a1,a7]{Miki Nakajima}
	\ead{mnakajima@rochester.edu}
\author[a5]{Erik Asphaug}
	\ead{easphaug@gmail.com}
\address[a1]{Department of Physics and Astronomy, University of Rochester, Rochester, NY 14627, USA}
\address[a2]{Department of Computer Science, University of Rochester, Rochester, NY 14627, USA}
\address[a3]{Colorado Center for Astrodynamics Research, The University of Colorado Boulder, UCB 431, Boulder, CO 80309-0431, United States}
\address[a4]{Department of Geology and Geological Engineering, Colorado School of Mines, Golden, CO 80401, USA}
\address[a5]{Lunar and Planetary Lab, University of Arizona, Tucson, AZ, USA}
\address[a6]{Laboratoire Lagrange, Universit\'e C\^ote d'Azur, Observatoire de la C\^ote d'Azur, CNRS, C.S. 34229, 06304 Nice Cedex 4, France}
\address[a7]{Department of Earth and Environmental Sciences, University of Rochester, Rochester, NY 14627, USA}
\cortext[cor1]{Corresponding author, \texttt{ewrig15@ur.rochester.edu}}

\begin{document}

\begin{abstract}
We consider how an impact generated  seismic pulse affects the surface of an asteroid distant from the impact site.
With laboratory experiments on dry polydisperse gravel mixtures, we track the trajectories of  
particles ejected from the surface by a single strong upward propagating pressure pulse. 
High speed video images show that ejecta trajectories are independent
of particle size, and collisions primarily take place upon landing.
When they land particles are ballistically sorted, as proposed by \citet{shinbrot17},
leaving larger particles on the surface and smaller particles
more widely dispersed.     A single strong pulse can leave previously buried boulders stranded on the surface.
Boulder stranding due to an impact excited seismic pulse is an additional mechanism that could leave 
 large boulders present on the surface of rubble asteroids such as 162173 Ryugu, 101955 Bennu and 25143 Itokawa.
\end{abstract}

\maketitle

%\linenumbers

\section{Introduction}

Particle size segregation on rubble covered asteroids such as 25143 Itokawa and the presence of large
boulders on the surface are usually explained with  
 the Brazil nut effect (BNE) \citep{matsumura14,perera16}, by which vibrations 
in the presence of gravity slowly propel larger particles to the surface. 
In the case of 25143 Itokawa, the size distribution seems to be correlated to the total gravitational potential with large and small boulders in regions of high and low potential respectively \citep{tancredi15,miyamoto07}.
The Brazil nut effect brings the largest nuts in a shaken bowl of mixed nuts to the top (e.g., \citealt{rosato87}),
and is responsible for the appearance of boulders in agricultural fields after repeated cycles of frost heave.
The Brazil nut effect is mediated through a number of possible mechanisms (e.g. ratcheting, convection), all dependent on gravitational acceleration.
Smaller particles percolate or slip beneath larger ones and ratchet the largest ones upward 
 \citep{williams76,rosato87,jullien92,hong01,maurel17}. 
Alternatively a convection pattern 
%brings larger particles to the surface but prevents these same 
dredges particles up to the surface, but inhibits larger
particles from sinking \citep{knight93,chujo18,matsumura14}.

Impact induced seismicity is  important on small asteroids due to their
low surface gravity and small volume which limits vibrational energy dispersal 
\citep{cintala78,cheng02,richardson04}.
Seismic disturbances can destabilize loose material resting on slopes, causing landslides
%downhill flows 
\citep{titley66}, and crater degradation and crater erasure
\citep{richardson04,thomas05,richardson05,asphaug08,yamada16}.
Regions of different crater densities on asteroid 433 Eros can be explained by strong impacts that 
 erase craters \citep{thomas05}.

Unfortunately, little is currently known about how impact generated seismic waves are excited, 
dispersed, attenuated and scattered in asteroids.  The rapidly attenuated seismic pulse or `jolt' model  \citep{nolan92,greenberg94,greenberg96,thomas05} is
consistent with strong attenuation in laboratory granular materials at kHz frequencies \citep{odonovan16}, but
qualitatively differs from
the slowly attenuating seismic reverberation model \citep{cintala78,cheng02,richardson04,richardson05}, that is supported
by measurements of slow seismic attenuation rates in lunar regolith \citep{dainty74,toksoz74,nakamura76}.
While both impact induced seismic `jolt' 
%\citep{nolan92,greenberg94,greenberg96} 
and reverberation processes can cause crater erasure and rim degradation \citep{richardson04,richardson05,thomas05,asphaug08}, 
it is often assumed that size segregation induced by the Brazil nut effect depends on sustained vibrations or reverberation  
(e.g., \citealt{miyamoto07,tancredi12,matsumura14,tancredi15,perera16,maurel17,chujo18}).
However, a series of strong taps, separated in time,
can also bring larger particles to the surface (e.g., \url{https://www.youtube.com/watch?v=XTM-okBCX8U}).

Even cohesionless or low strength rubble would transmit an impact generated compression wave. 
However,  asteroid rubble may have a low tensile strength of $\sim 10 $ Pa \citep{sanchez14,scheeres18}.  
Impact generated seismic waves
may not effectively rebound or reflect from asteroid surfaces.
When they reach the surface, strong  seismic waves can cause the
surface to deform, induce landslides and loft particles off the surface (e.g., \citealt{tancredi12}), reducing
the amplitude of reflected waves, increasing the attenuation rate 
and reducing the seismic reverberation time.  

Simulations have shown that the Brazil nut effect is still effective in low gravity environments with $g\sim10^{-4}$ \citep{tancredi12,chujo18,matsumura14,maurel17}. Simulations by \citeauthor{matsumura14} show that the time for a particle to rise to the top of a granular mix is proportional to the square root of the gravitational acceleration of the body. The low gravity of asteroids will only affect the time scale of the granular flow due to the Brazil nut effect \citep{maurel17}. That is, size segregation due to the BNE is still effective in low gravity environments, it only takes longer for the process to proceed.

%may not be effective in low surface  gravity \citep{maurel17} 
%compared to the laboratory  (e.g, \citealt{williams76,chujo18}).

\citet{shinbrot17} proposed that pebbles accreting or falling onto the surface of an  
asteroid would tend to rebound from boulders, 
but sink into pebbly regions  leading to lateral size segregation on a surface, a process they
called `ballistic sorting'.  Since a pebble sea contains numerous pebbles, a particle landing there 
causes numerous collisions,  making granular beds good impact absorbers 
due to their low coefficient of restitution \citep{meakin86,jenkins97}.  
\citet{shinbrot17} illustrated the effect with numerical simulations and by experimentally dropping
glass beads and pebbles onto flat surfaces and surfaces containing mixtures of pebbles. 
They did not discuss the source or velocity distribution of the landing material.

Using  numerical simulations, \citet{tancredi12} explored launching of ejecta from a surface  in low gravity due to a 
subsurface seismic pulse traveling through a granular medium.   
However, \citet{tancredi12} did not study the kinematics of the launched ejecta during landing
%\citet{shinbrot17} and \cite{tancredi12} did not explore 
and whether ejecta launched by impact excited seismic waves could leave boulders stranded on the surface.
% and   they would have missed ballistic sorting.

Using laboratory experiments in a polydisperse granular medium, we explore surface modification due to ballistic sorting of the ejecta from an
upward propagating subsurface pressure pulse that is strong enough to launch surface particles into the air.
In section \ref{sec:exp} we describe our experimental setup.    With high speed video in section \ref{sec:trackpy} we 
use particle image velocimetry to measure velocity vectors of ejecta particles.
%Our experiments illustrate that a single pressure pulse can strand previously embedded
%boulders on the surface. 
%The particle image velocimetry points to the ballistic sorting process \citep{shinbrot17} as the likely mechanism.
In section \ref{sec:asteroid} we use scaling relations for seismic energy efficiency of an impact to estimate
the importance of  impact generated seismic pulses as a size segregation and surface modification process 
on rubble asteroids.  A summary and discussion follows in section \ref{sec:sum}.

\section{Laboratory experiments of upward propagating pressure pulses in a polydisperse granular medium}
\label{sec:exp}

%\subsection{Experimental setup}
%\ref{sec:setup}

\begin{figure}
\centering\includegraphics[width=0.5\textwidth]{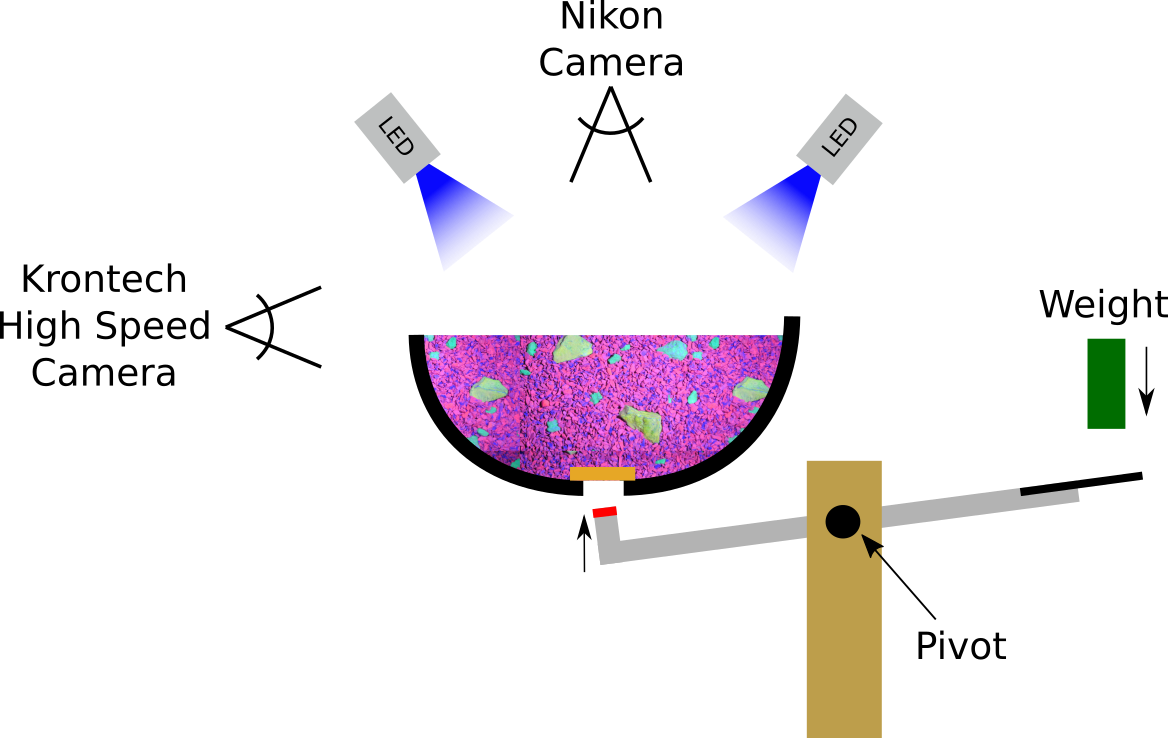} 
\caption{Illustration of our experimental setup for filming particles launched into the air by a subsurface pulse. A lever arm (gray bar) rotates on a pivot when a weight is dropped onto a metal plate (black line at end of lever). A metal piece attached to the lever rotates up and hits a metal disk inside the bowl of gravel (orange rectangle at bottom of bowl). The bowl's motion is reduced  by adding weights to the rim (not shown). Two blue LEDs are used to light gravel particles that are painted with pink, green, and yellow fluorescent paints. The largest particles are painted yellow, medium sized particles are green, and smallest particles are in pink. The gravel illustrated in the bowl is not to scale. A Nikon camera was positioned above the bowl to film the ejecta from above. A Krontech high speed camera was positioned at the side of the bowl and at a height approximately level with the surface of the gravel. For all experiments the gravel had a depth of 6 cm at the center. The lever arm allowed us to carry out experiments with similar
pulse amplitudes.  The pulse amplitude was varied by changing the mass of the dropped weight or its initial height.  The pulse duration was adjusted by adding or removing a small piece of foam to the tip of the impactor (shown as a red square at the tip of impactor). 
}
\label{fig:arm}
\end{figure}

A bowl of granular polydisperse colored dry gravel is used to mimic the uppermost layer of a rubble pile asteroid. We hit the bottom of the gravel (not the bowl itself) with a strong impact and film the granular surface with high speed video as particles are launched into the air.  
An accelerometer in the bowl is used
to characterize the pressure pulse strength and duration as a function of time. 
The experimental setup is illustrated in Figure \ref{fig:arm}.

Soft sphere granular models find that wave speed depend on pressure as this sets the strength of elastic contacts in force chains \citep{makse04}.
With lower speeds near the surface, seismic waves would approach the surface at a near normal angle.   To mimic this behavior we restrict our study
to upward directed impacts into a granular media with an initially horizontal upper surface and in a container with a horizontal flat base.
We have sifted the gravel to remove small particles (sand) and minimize the dissipative role of air flow. 
We recognize that the modest size of our experiments,  the fact that we carry them out in air, not vacuum, and that our lab is not in milli or microgravity, may give us experimental  results that are not representative of seismicity on a rubble pile asteroid. 

The granular medium consists of materials commonly used in construction in Upstate NY.
We use small and larger particles of crushed shale and medium sized particles of quartzite.  
%To minimize the dissipative role of air flow in small voids and drag, we sifted the medium to remove particles with widths smaller than 1.5 mm. 
Since our gravel is irregularly shaped all sizes were measured along their longest dimension.
The smallest particles are 5 to 10 mm in length. 
The medium sized particles range in size from 15 to 25 mm and the largest particles from 30 to 55 mm. 
We mixed the gravel by hand.
The mixture is predominantly the small particles, with about 20 medium sized particles, and up to five of the largest particles.

To minimize the role of air we sifted out powdered material and particles finer than about 5 mm. 
Experiments have shown that a fluid (e.g. air, water) present in a granular system can accelerate the ratcheting of a large particle compared to just convectively driven motion \citep{clement10,naylor03}.
\citet{mobius05} showed that if the relative sizes of particles in the mix is small ($D/d<10$) then air-driven size separation is negligible. They also found that air drag on the intruder was negligible and only the drag from the smaller particles contribute significantly.
For the size distributions in our mix we find the relative size for the medium to small particles to be 2.5 - 3 and the relative size for the large to small particles to be 5.5 - 6.
Therefore, the effects of air can be safely ignored in our experiments.

Gravel particles are painted with different fluorescent colors so we can see if particle velocities are dependent on particle size. 
The paint was applied from a spray can in a thin layer to not change the shape of the particles.
We measured the angle of repose to be 34 degrees for both the bare and painted gravel mixture.  
%34 degrees and found it to be the same for painted and bare particles.  
Painting the rocks did not significantly change the friction between particles.

The experiment was lit with bright blue LEDs, causing the painted gravel to fluoresce.
This allowed the tracking of different sized particles so we could see if ejecta velocities and particle trajectories were  dependent on particle size.
Smaller particles are painted fluorescent orange, though the un-dyed regions appear blue in our images because of our lighting.
Medium sized particles are painted fluorescent green and the largest ones are painted fluorescent yellow.
We were careful to use fluorescent paints that were detected as bright by our high speed camera, 
finding that our camera is insensitive to common red fluorescent paints and that different shades of blue paints were indistinguishable. 
%The experimental setup is illustrated in Figure \ref{fig:arm}.

The gravel was held in a 25 cm diameter plastic bowl with a 5 cm diameter hole drilled in the bottom. 
A flat metal disk with a diameter of 8 cm was placed over the hole to spread the pulse over a wider area. 
The disk was not fastened to the bowl so free to move, and it is strong enough not to deform under the weight of the gravel or when it is struck by the lever arm.
This contributed to the impactor's momentum going into accelerating the gravel, not the bowl.
We covered the disk and the inside of the bowl loosely with black cloth wrap to keep granular material from falling below the disk and through the hole. 
Since the disk was not held in place and loosely covered with a cloth air can still flow freely in the granular mix.

Our granular mixture filled the bowl to a depth of about 6 cm at the center.
Five 1 kg weights are placed symmetrically on the bowl rim to reduce elastic vibrations in the bowl and reduce how far the bowl moves during the impacts. 
We painted a  $1\times 0.5$ cm scale bar on one of the wieghts in green fluorescent paint.
We use this scale bar to derive the pixel scale used to track particle velocities and motions between frames. More details  on particle tracking are given in section \ref{sec:trackpy}.

%We impacted the metal disk (not the bowl itself) with a levered hammer so
%as to excite a pressure pulse in the granular medium. 
A weight up to one kilogram was dropped from a height of 15 cm onto a metal plate at the back of the lever arm. 
The arm then rotates and the hammer end strikes the metal disk inside the bowl. 
The levered hammer provided repeatable single pulses. 
The hammer stays in contact with the metal disk until the weight is manually removed from the plate.
The pressure pulse travels through the granular media upward and emerges on the top surface as a radial wavefront where it lofts gravel into the air. After each impact the gravel is remixed as specified above.

We filmed the ejecta from the side with a Chronos 1.4 high speed camera at 1057 frames per second and from above at 30 frames per second with a Nikon V1 for all experiments.
An accelerometer was %(the ADXL377 chip),
buried in the gravel about 3 cm below the surface, allowed us to record acceleration as a function of time in the gravel. 
%Data from the accelerometer was recorded using the analog to digital converter on an Arduino Mega.
Figure \ref{fig:accel} shows the accelerometer data for each experiment listed in Table \ref{tab:exp}.
The Arduino Mega was programmed to record data when the accelerometer readings exceeded about 8g. This threshold is given by the dashed black line in theses figures.
Experiments are labelled consecutively by number in our lab notebooks.  The experiments discussed in this paper are denoted with numbers 39, 41, 42 and 43.  

Only four experiments are shown in this paper because these were the best quality complete data sets. A complete data set included high quality high-speed (1057fps) and normal speed (60fps) videos, and accelerometer measurements. High quality videos were considered those that had the ejecta curtain entirely in focus and in frame, good lighting so that the ejecta curtain would not cast a shadow on itself. Quality accelerometer data did not include clipping in the data from the accelerometer being too slow to capture it.
An experiment's data set was considered incomplete if the quality of the videos or accelerometer did not meet requirements, or if the cameras or accelerometer were not triggered resulting in no data. We observed ballistic sorting upon material landing for all experiments performed regardless of data acquired.
%Experiment 40 was discarded because the accelerometer data was clipped and so unreliable. Experiments before 39 were not useable for a combination of reasons (e.g. poor lighting, videos did not record, unfocused, etc.).

An impact can be described with a maximum acceleration value and a pulse duration.
We can adjust the pressure pulse amplitude, by dropping a heavier weight or increasing its initial height.  
We adjust the pulse duration by inserting or removing a piece of foam between the impactor and plate. 
We see broader pulses with thicker foam.

\subsection{Boulder stranding with a single impulse}

\begin{table}
%\vbox to80mm{\vfil
\caption{\large  Impact  Experiments \label{tab:exp}}
\begin{tabular}{@{}lllllll}
\hline
Experiment \#  & Peak Acceleration  &  Pulse Duration\\
\hline
39 &  12 g  &  10 ms\\
41 & 18.5 g  &  10 ms\\
42 &  50 g  &  3 ms\\
43 &  37 g  &  3 ms\\
%Ratio of  & 7  \\
\hline
\end{tabular}
{\\ Notes: The experimental setup is shown  in Fig. \ref{fig:arm}.  Peak acceleration are the maximum seen
in the acceleration profiles (see Figure \ref{fig:accel}).
Pulse durations were measured from the full duration at half maximum of the acceleration profile.
} %}
\end{table}

\begin{figure*}
\centering\includegraphics[width=7in]{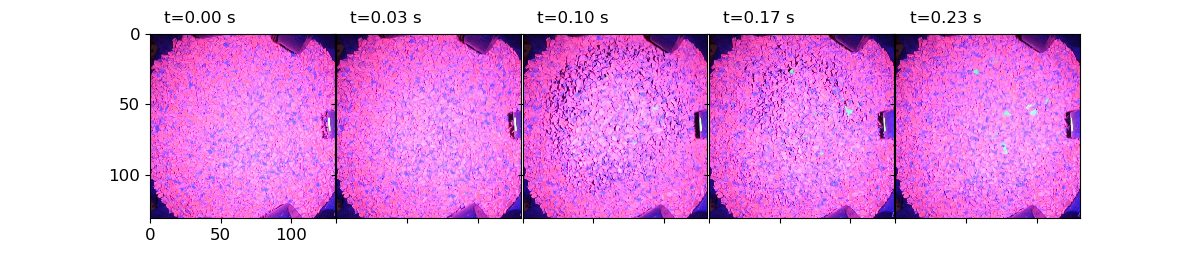} 
\centering\includegraphics[width=7in]{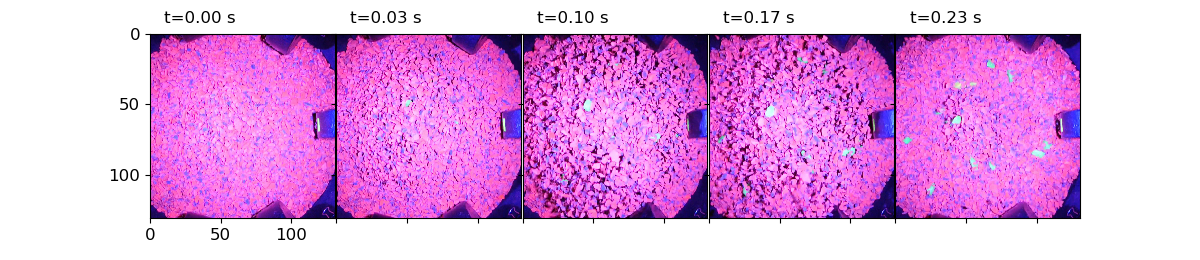} 
\centering\includegraphics[width=7in]{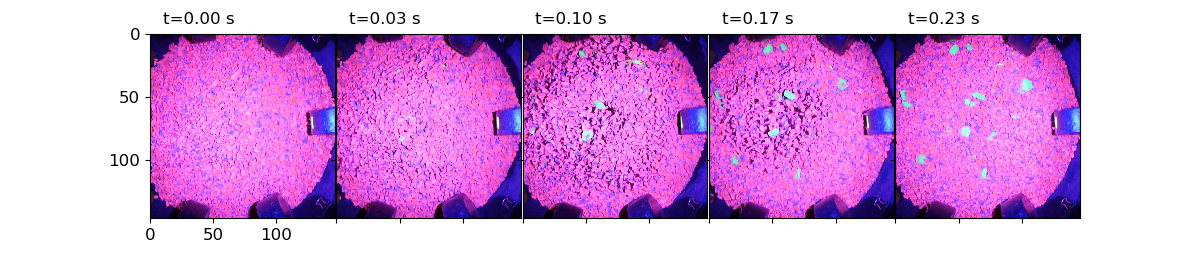} 
\centering\includegraphics[width=7in]{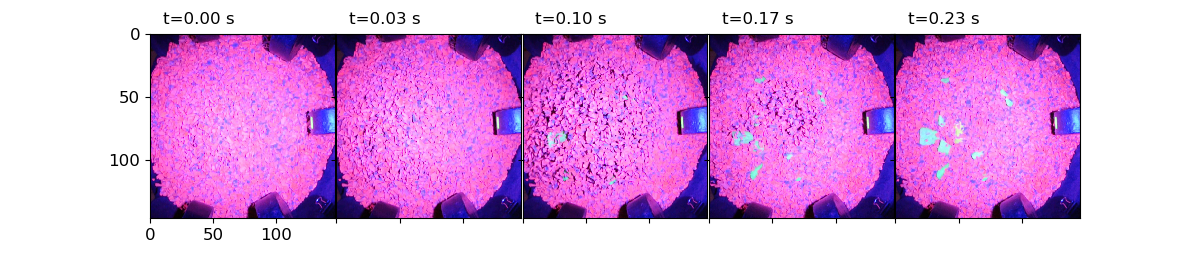} 
\caption{Frames from regular speed video viewing the granular surface from above and looking directly downward.  
The top two rows show the weaker pulse (39 and 41 respectively) impact experiments and the two bottom rows show the
stronger pulse impact experiments (42 and 43 respectively). These experiments are listed in Table \ref{tab:exp}. 
The $x$ and $y$ axes for leftmost panels give the scale in mm.   We measure time from when surface particles
start to move upward.  The time of each image is listed on top of each frame. 
Particles are painted with fluorescent paints and are lit with blue LEDs.
 Larger particles are painted green and yellow
and smaller particles are painted pink. Some sides of the smaller particles are not covered in paint and appear blue due to the lighting.
Frames on the right show that larger particles are lofted due to the subsurface impulse and left stranded on the surface afterwards.
Previously buried boulders are stranded on the surface in all our experiments. 
%Experiments with shorter pulses did not result in fewer boulders being stranded on the surface.
}
\label{fig:posts}
\end{figure*}

During each impact experiment we take two simultaneous videos, one from above (shown in Figure \ref{fig:posts}) and one from the side
(shown in Figures \ref{fig:hs39} - \ref{fig:hs43}).
Thumbnails (and links) for the eight videos are shown in Figure \ref{fig:stills}.
Simultaneously we measure acceleration as a function of time with the accelerometer (accelerometer profiles are shown in Figure \ref{fig:accel}). 
The pulse duration was measured by taking the full duration at half maximum of the acceleration profile.
The impact experiments we discuss are listed in Table \ref{tab:exp}.
In the table, we list peak acceleration values in units of g, the gravitational acceleration on Earth, 
and the pulse duration in milliseconds measured from the accelerometer profiles. 

The pulse was measured using an ADXL377 chip 200g accelerometer and recorded using the analog to digital converter of an Arduino Mega. 
%The accelerometer was located about 3 cm below the surface of the bowl. 
We found that the acceleration profiles in time are not strongly dependent upon the depth that the
accelerometer was buried, unless it was touching the base plate or surface. 
Figure \ref{fig:accel} shows the pulse profiles (acceleration as a function of time) recorded for each experiment. 
Pulses for experiments 39 and 41 had amplitudes less than 20g and a duration of 10ms. 
Experiments 42 and 43 had pulses with amplitudes greater than 30g and duration of 3ms.
The time for sound waves to travel across the depth of the bowl (0.6 ms if at a speed 100 m/s across 6 cm) 
is less than the durations of our pulses, so these durations give the durations of bulk material motions in the bowl. 

Figure \ref{fig:posts} shows images taken from the 30 frames per second video looking down on the surface at different times for
the four impact experiments listed in Table \ref{tab:exp}.
From top to bottom, each panel corresponds to experiments 39, 41, 42, 43 respectively.
We measure time from when the surface starts to move.
In all impact experiments, we see large particles, seen as green or yellow in the images, 
are buried prior to impact and are on the surface after impact. 
The large particles initially have random depths but we checked that large particles were not initially shallower than 0.5 cm below the surface. 
It is clear from the figures that regardless of different pulse amplitudes and durations, large particles are stranded on the surface after a single impact. The particles on the surface after impact are also seen in the high speed videos that are included in the supplemental material.
The experiments are robust and repeatable. 
The experiments show that a series of pulses, continuous vibration, or reverberation is not necessary to bring 
initially buried large particles to the surface.

We found that pulses with larger peak accelerations are able to excavate deeper boulders.
Large particles that are already on the surface tend to stay there after a second or third impact.
Pulses that are too weak (less than 10 g) and short (less than a few ms) were ineffective at stranding larger particles on the surface.
All the experiments shown here launched
particles  into the air, so they all have pulses with accelerations above surface gravity.
The ejecta travel upward a few cm and so travel a vertical distance that is greater than the stranded large particle heights.
The impacts cause all the material in the bowl to move upwards, so the depth of launched material
also exceeds the large particle lengths.  We will use these conditions  
when we discuss regimes for boulder stranding in section \ref{sec:asteroid}.

%Specific requirements must be met to strand a boulder on the surface.
%A single pulse must accelerate material above that of surface gravity, ejected material must travel a vertical distance that is greater than the boulder height, and the depth of the material ejected is larger than the boulder height.
%As long as there are large particles in the bowl we see some reach the surface after a single impulse. 

\subsection{Particle tracking}
\label{sec:trackpy}

We use the soft-matter particle tracking software package \texttt{trackpy} \citep{trackpy} to identify
and track gravel particles in the video frames.
\texttt{Trackpy} is a software package for finding blob-like features in video, tracking them through time, and analyzing their trajectories. It implements and extends the widely-used Crocker-Grier algorithm \citep{crocker96} in Python.

We measure velocity vectors at six different times in the high speed videos and for our four impact experiments.
At each start time we extract 20 consecutive frames and track particles in them, constructing about 1000 particle trajectories.
The particle velocities are computed from particle positions over four frames or 3.78 ms apart.
The velocity vectors are shown along with images from the high speed videos in Figures \ref{fig:hs39}-\ref{fig:hs43} for
the four impacts.
As before, times are labeled on the top of each image panel with time measured from when surface particles start to move upward.

In all four experiments, Figures \ref{fig:hs39}-\ref{fig:hs43} show that at early times the particles are moving upward as particles leave the surface. 
Ejected particles move together. 
During the time the gravel is launched until just before landing, nearby particles have similar velocity vectors. 
The last panels in Figures \ref{fig:hs39}-\ref{fig:hs43} show that as particles land the velocity vectors are randomly oriented, implying that scattering is taking place.  

\begin{figure}
\centering\includegraphics[width=\textwidth]{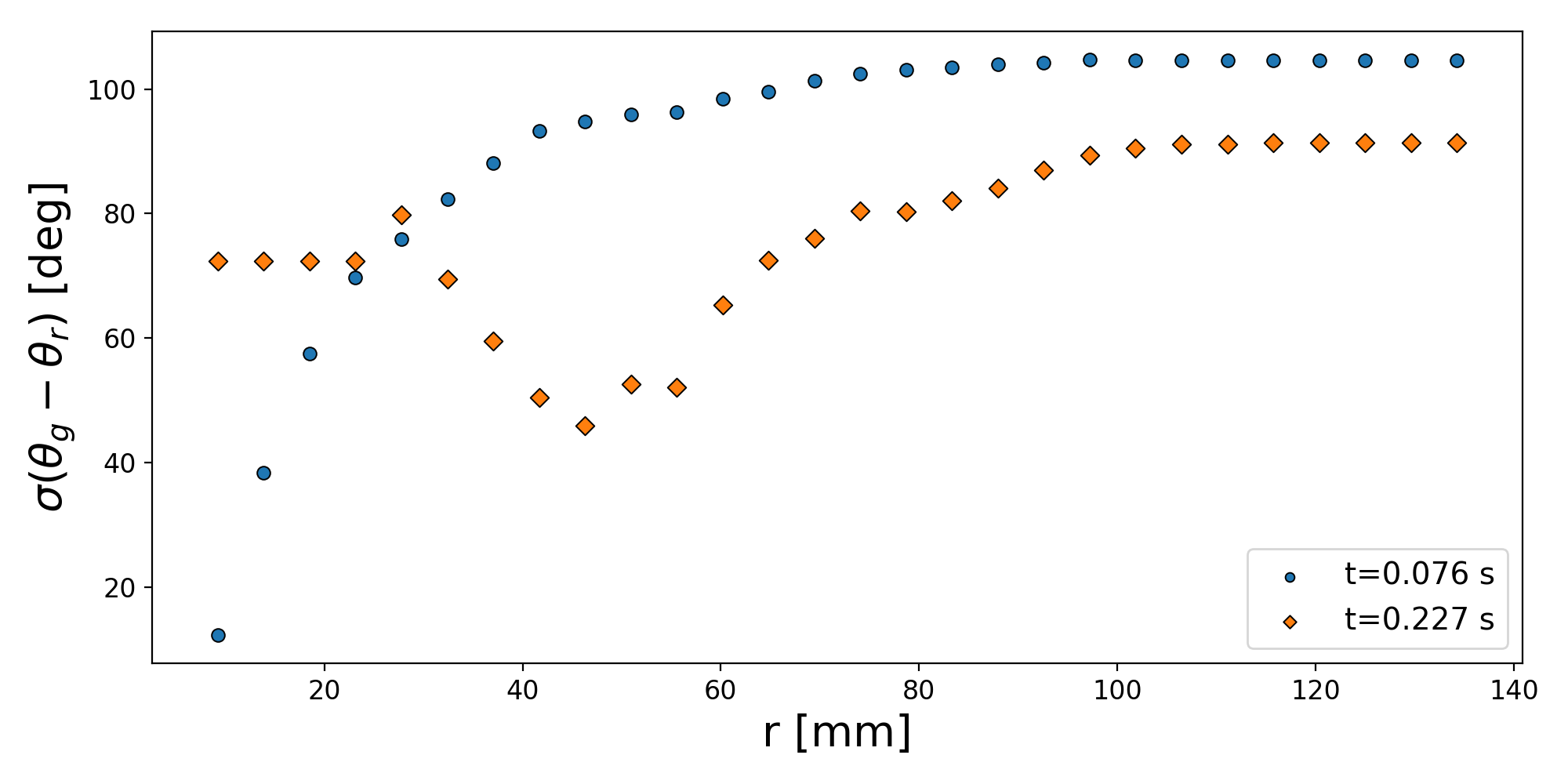} 
\caption{Plot of the standard deviation of the difference in the angles of the velocity vectors of the green and red particles as a function of the distance between a green particle and a red particle. The angles of all the red particles' velocity vectors within a distance r are subtracted from the angle of one green particle in the ejecta curtain. The blue circles show the standard deviation calculated for the ejecta at a time when the curtain reached near max height and the orange diamonds at a time when the material is landing. Both times used the same green particle. For earlier times, when the ejecta is moving upward, the standard deviation of the difference in velocity directions is small for nearby particles and increases as more distant particles are included. Later in time the standard deviation is large for nearby particles and varies with increasing distance. This shows that red particles in close proximity to the green particle are moving together with similar velocities early in time and are scattering later in time.
}
\label{fig:std_ang}
\end{figure}

\begin{figure*}
%\centering\includegraphics[width=3in]{39_a.png}  % this block is for two column layout
%\centering\includegraphics[width=3in]{39_b.png} 
%\centering\includegraphics[width=3in]{39_c.png} 
%\centering\includegraphics[width=3in]{39_d.png} 
%\centering\includegraphics[width=3in]{39_e.png} 
%\centering\includegraphics[width=3in]{39_f.png} 
\centering\includegraphics[width=2.5in]{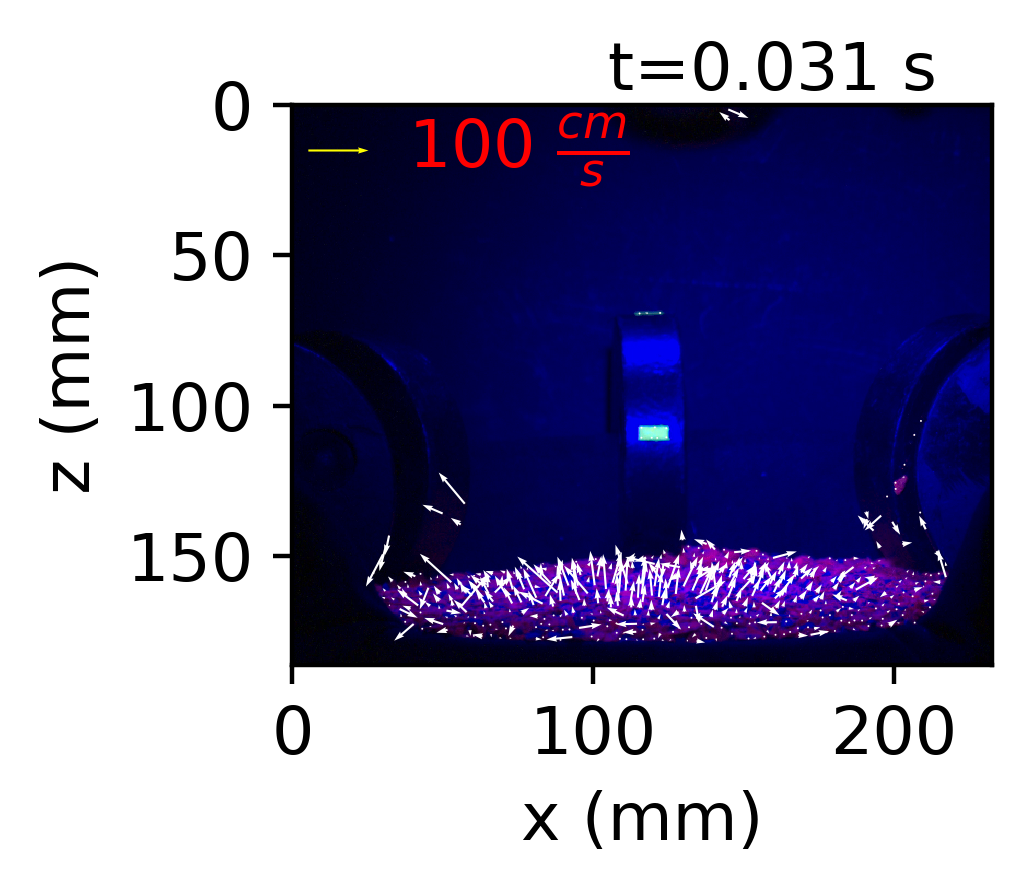} 
\centering\includegraphics[width=2.5in]{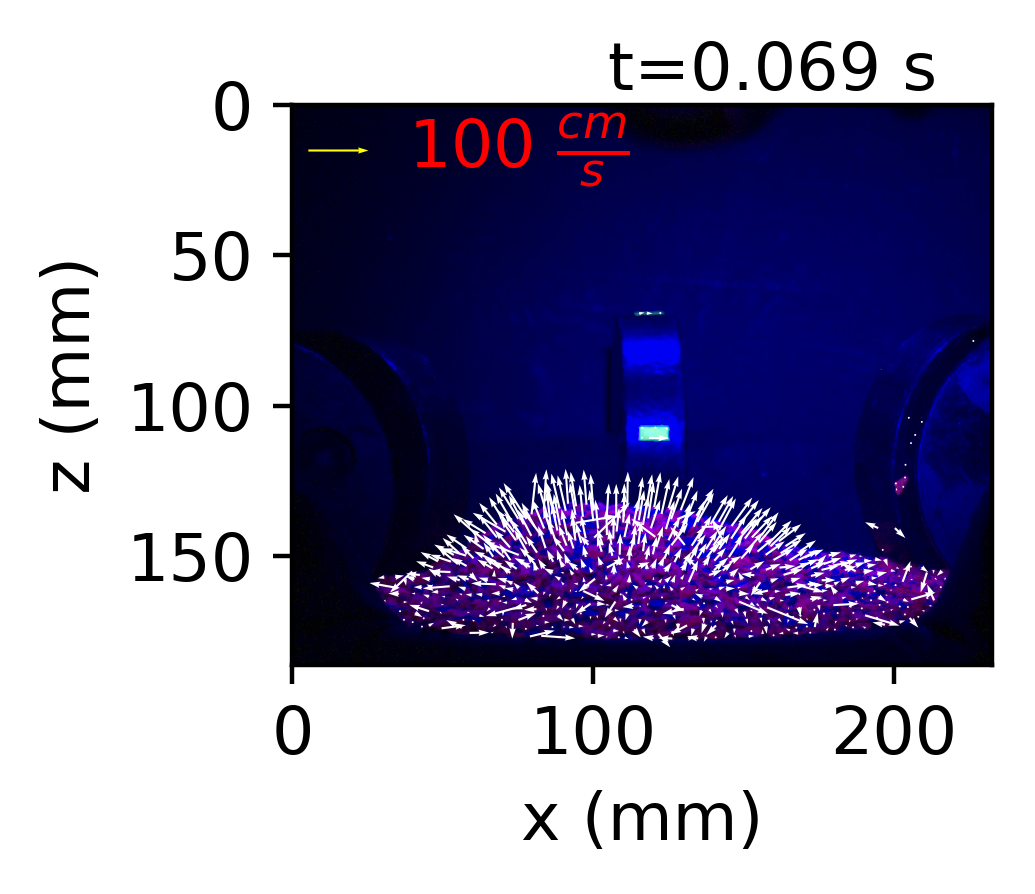} 
\centering\includegraphics[width=2.5in]{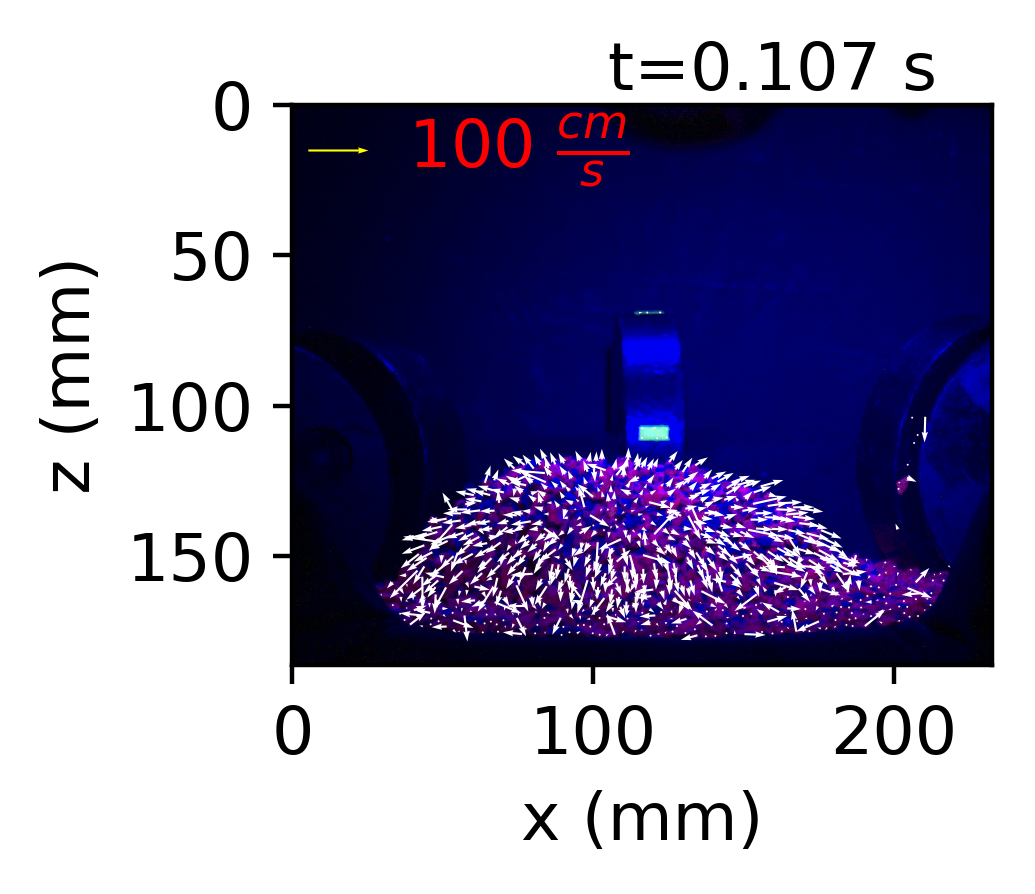} 
\centering\includegraphics[width=2.5in]{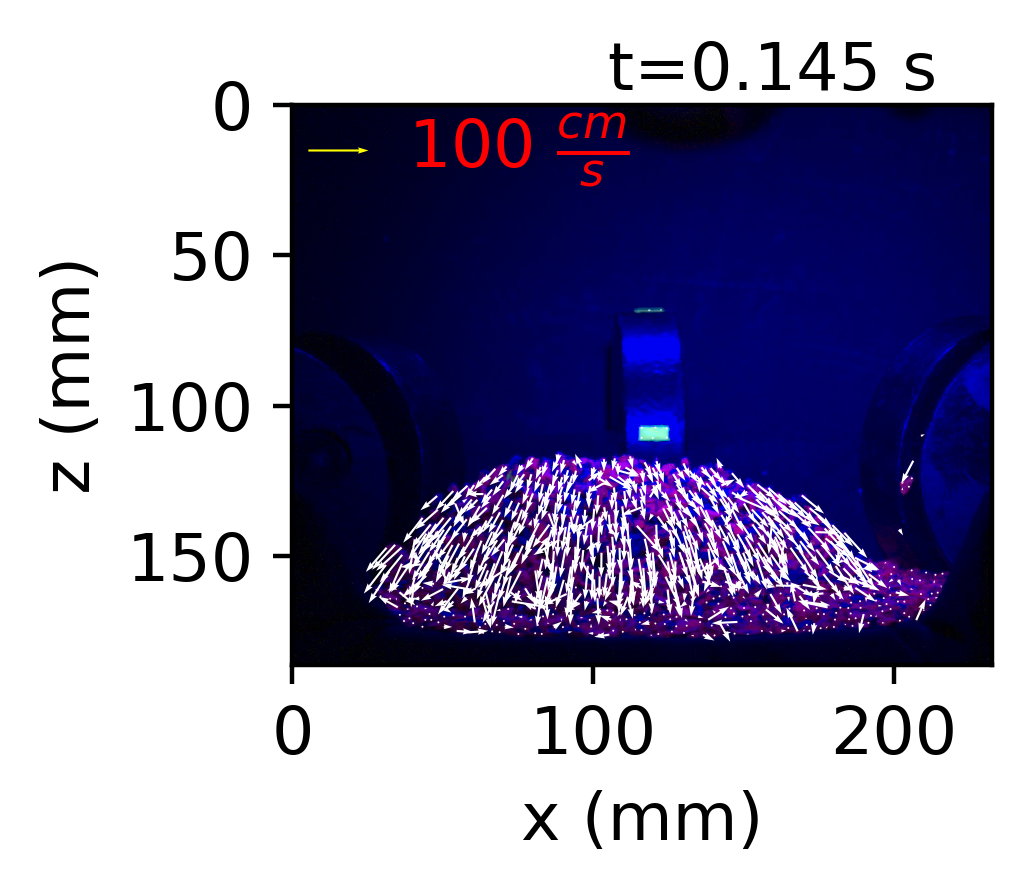} 
\centering\includegraphics[width=2.5in]{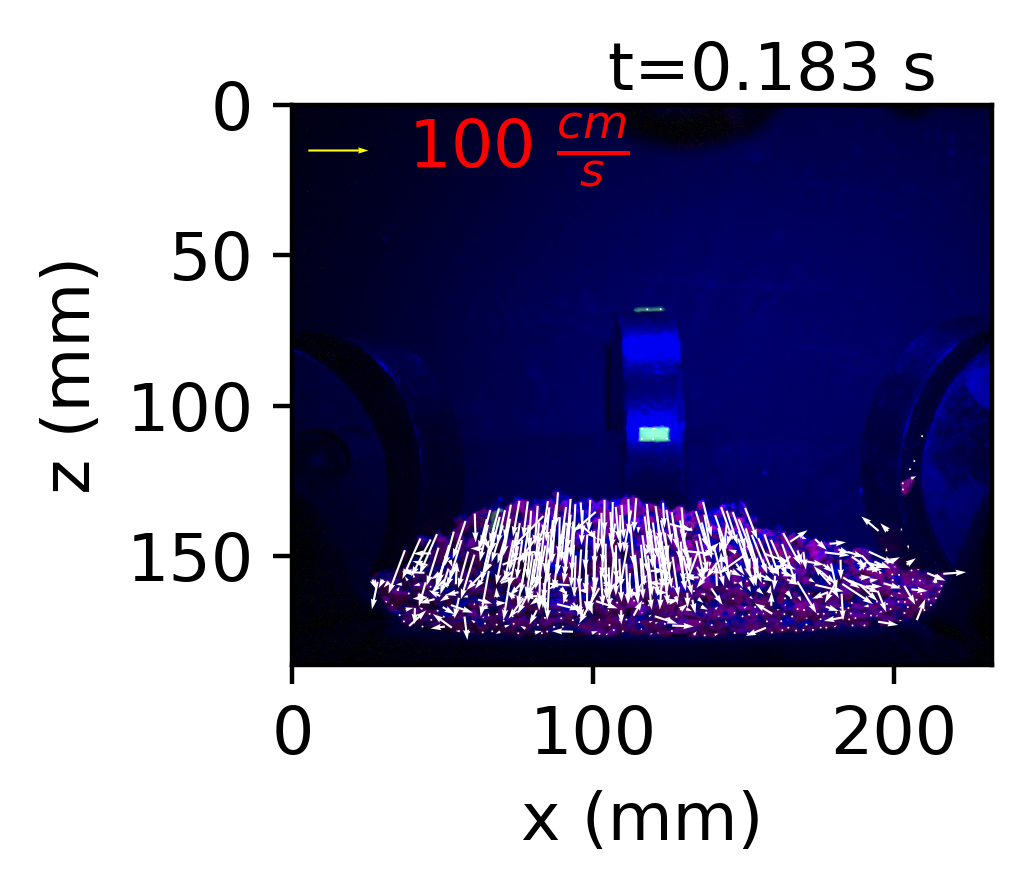} 
\centering\includegraphics[width=2.5in]{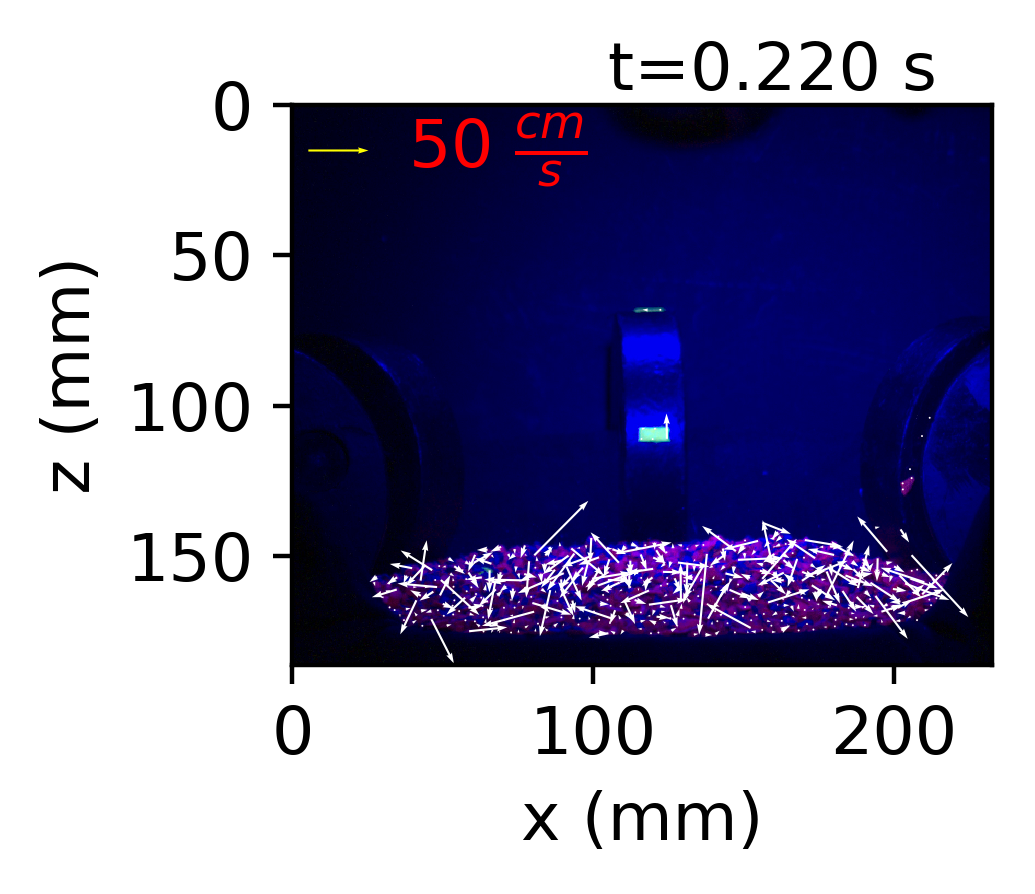} 
\caption{Snapshots from the high speed video (1057 frames per second) of experiment 39 overlaid with velocity vectors of individual particles.
The time of each panel is shown on the top of the panel and is  measured from when particles start to move upward. The yellow arrows in the top left corner show the scale of the velocity vectors in cm/s.
Nearby particles tend to have similar velocity  vectors throughout trajectory, except during landing when significant scattering occurs.
The blue rectangle in the background is a weight used to keep the bowl down.  On it, there is a green $1\times 0.5$ cm scale bar.
}
\label{fig:hs39}
\end{figure*}

\begin{figure*}
%\centering\includegraphics[width=3in]{41_a.png}  % this block is for two column layout
%\centering\includegraphics[width=3in]{41_b.png} 
%\centering\includegraphics[width=3in]{41_c.png} 
%\centering\includegraphics[width=3in]{41_d.png} 
%\centering\includegraphics[width=3in]{41_e.png} 
%\centering\includegraphics[width=3in]{41_f.png} 
\centering\includegraphics[width=2.5in]{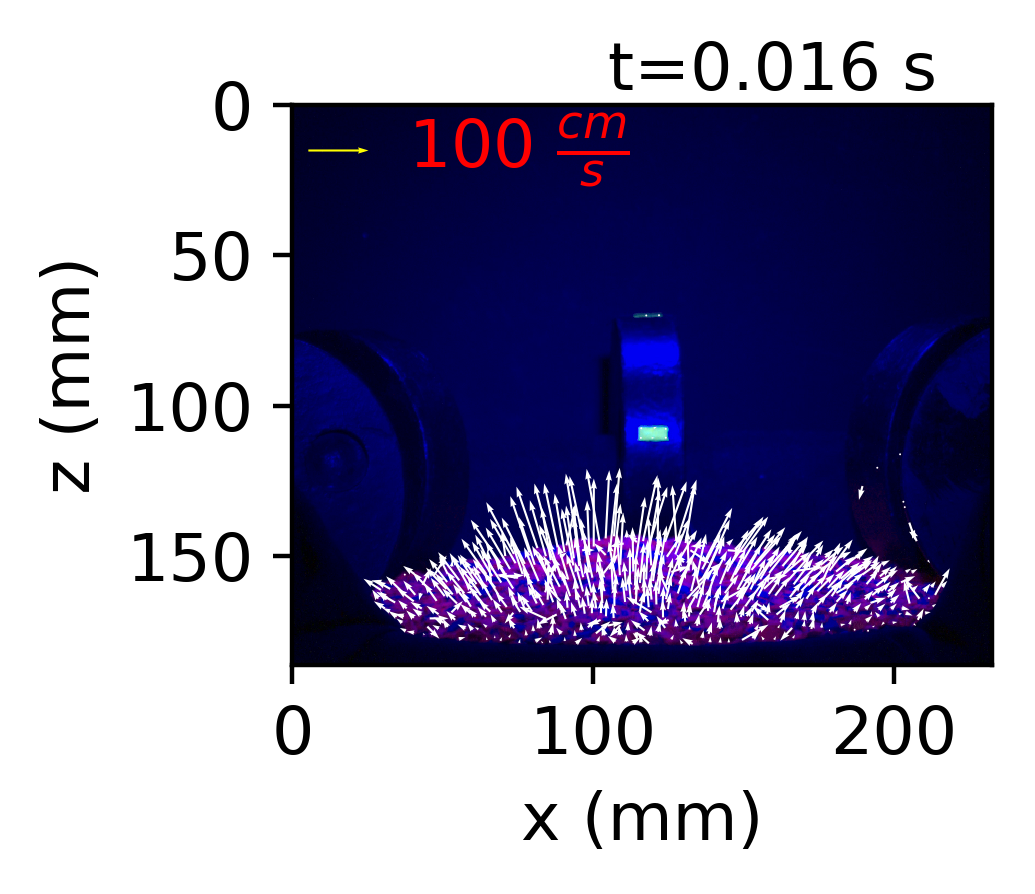} 
\centering\includegraphics[width=2.5in]{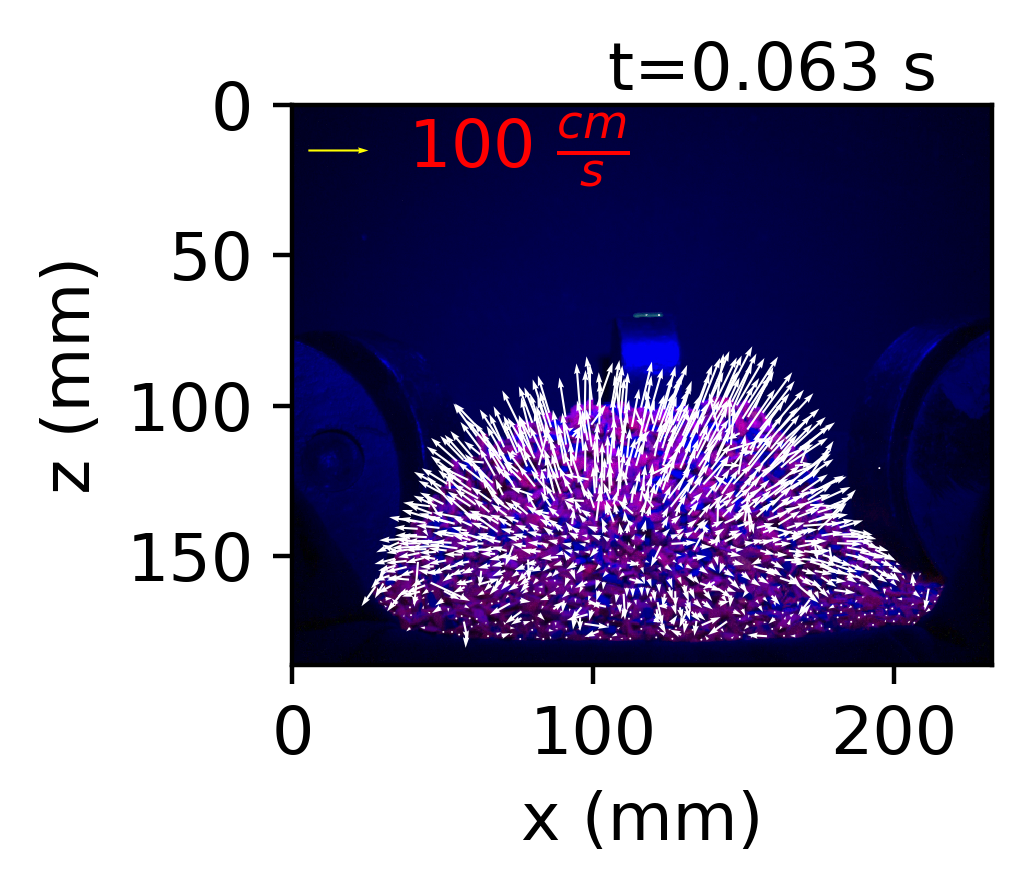} 
\centering\includegraphics[width=2.5in]{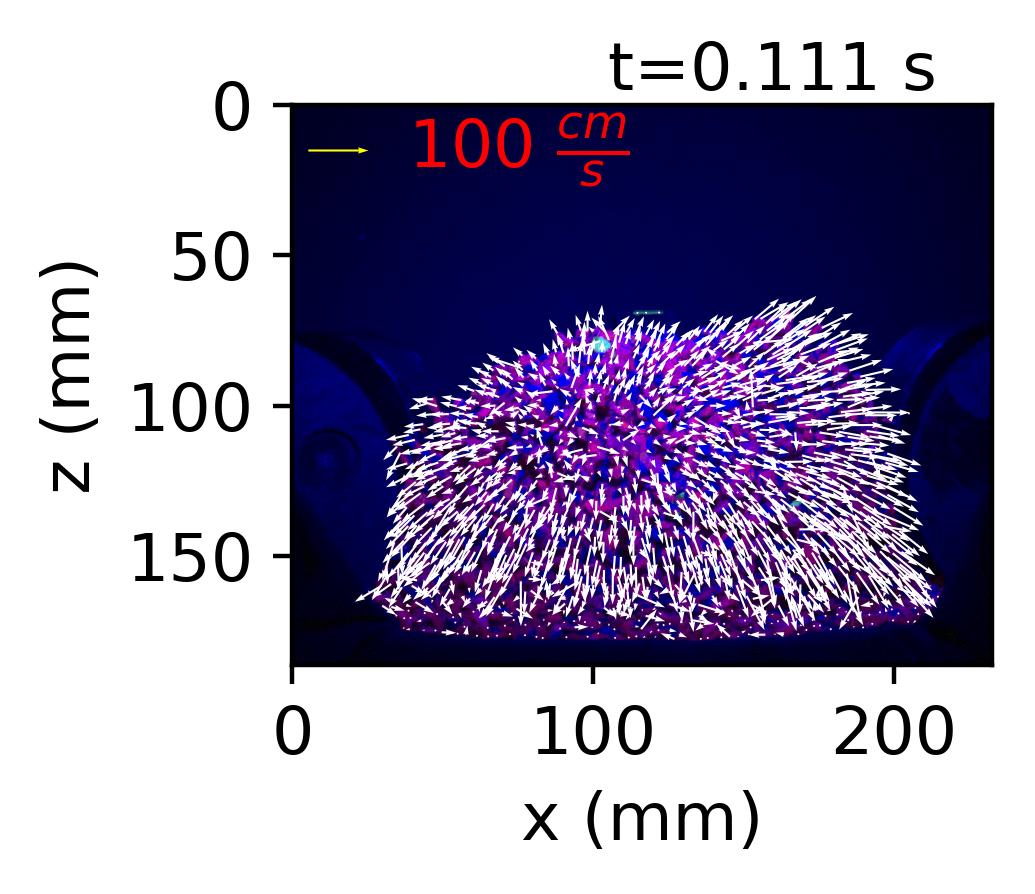} 
\centering\includegraphics[width=2.5in]{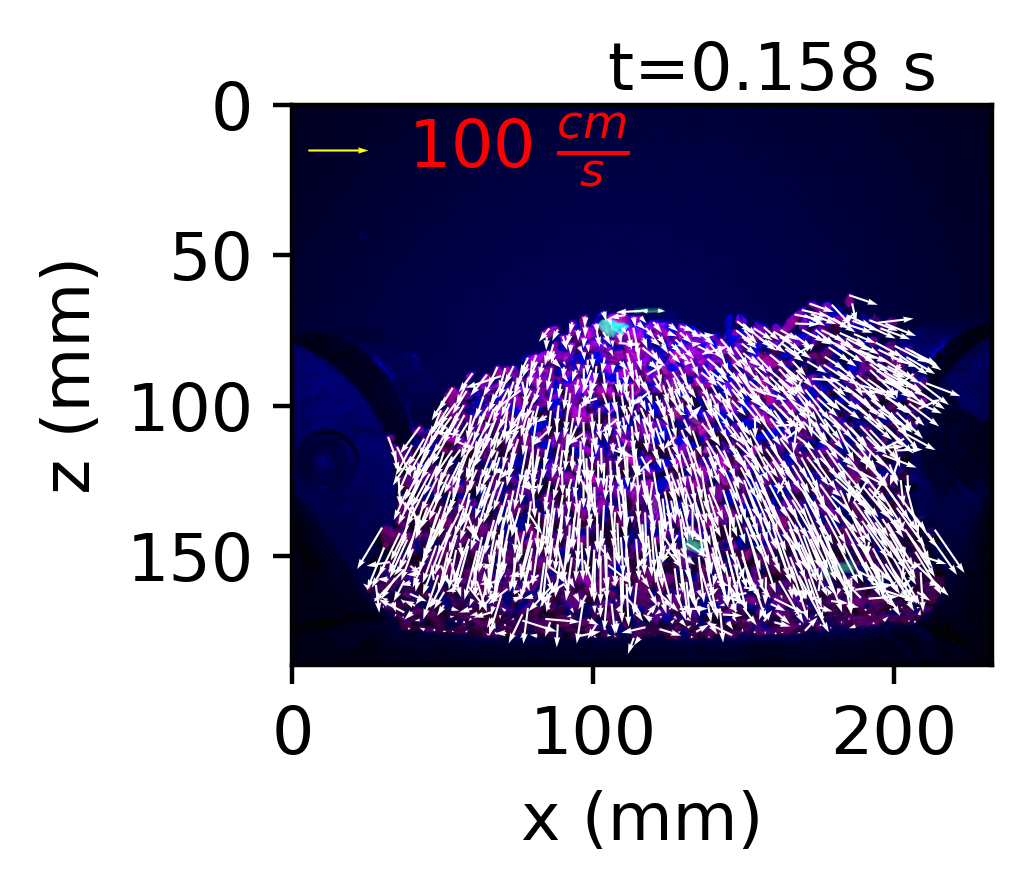} 
\centering\includegraphics[width=2.5in]{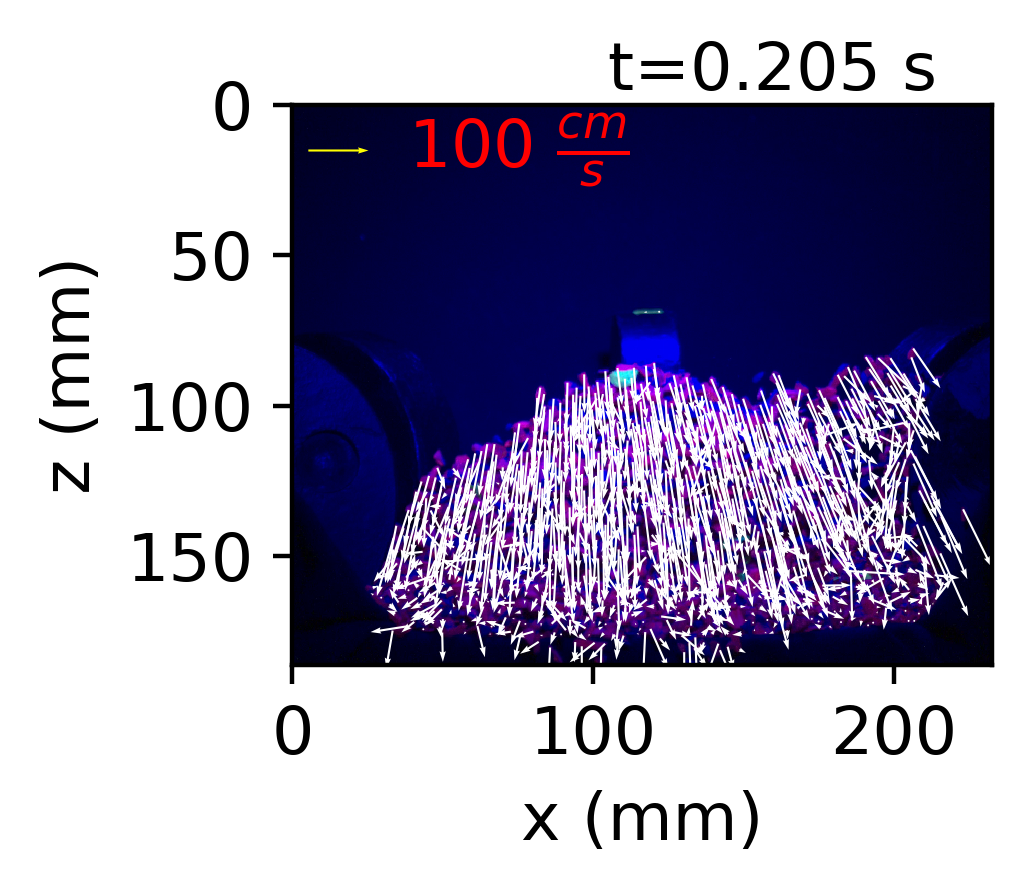} 
\centering\includegraphics[width=2.5in]{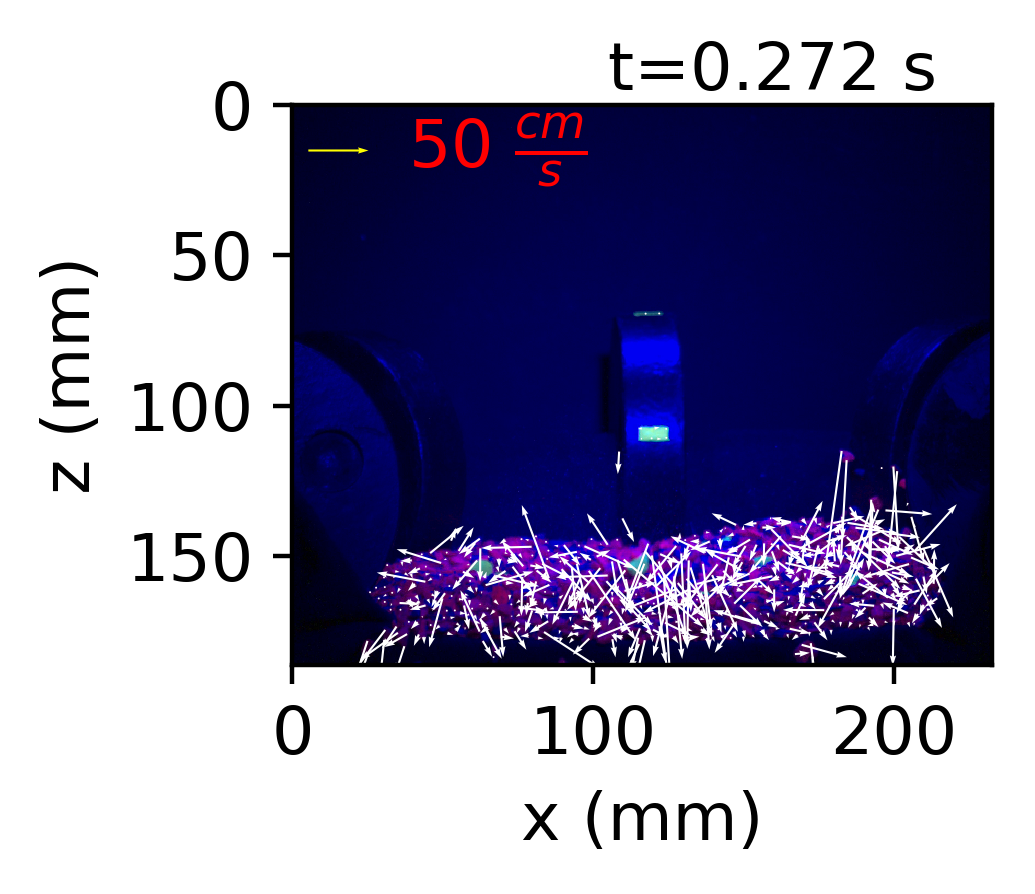} 
\caption{High speed video snapshots of experiment 41 and similar to Figure \ref{fig:hs39}. 
Snapshots show similar features as in Figure \ref{fig:hs39}.
A large green particle in the middle panels has similar velocity vectors to nearby smaller particles.
We see no evidence that larger particles have different velocity vectors than smaller particles prior to landing.
}
\label{fig:hs41}
\end{figure*}

\begin{figure*}
%\centering\includegraphics[width=3in]{42_a.png}  % this block is for two column layout
%\centering\includegraphics[width=3in]{42_b.png} 
%\centering\includegraphics[width=3in]{42_c.png} 
%\centering\includegraphics[width=3in]{42_d.png} 
%\centering\includegraphics[width=3in]{42_e.png} 
%\centering\includegraphics[width=3in]{42_f.png} 
\centering\includegraphics[width=2.5in]{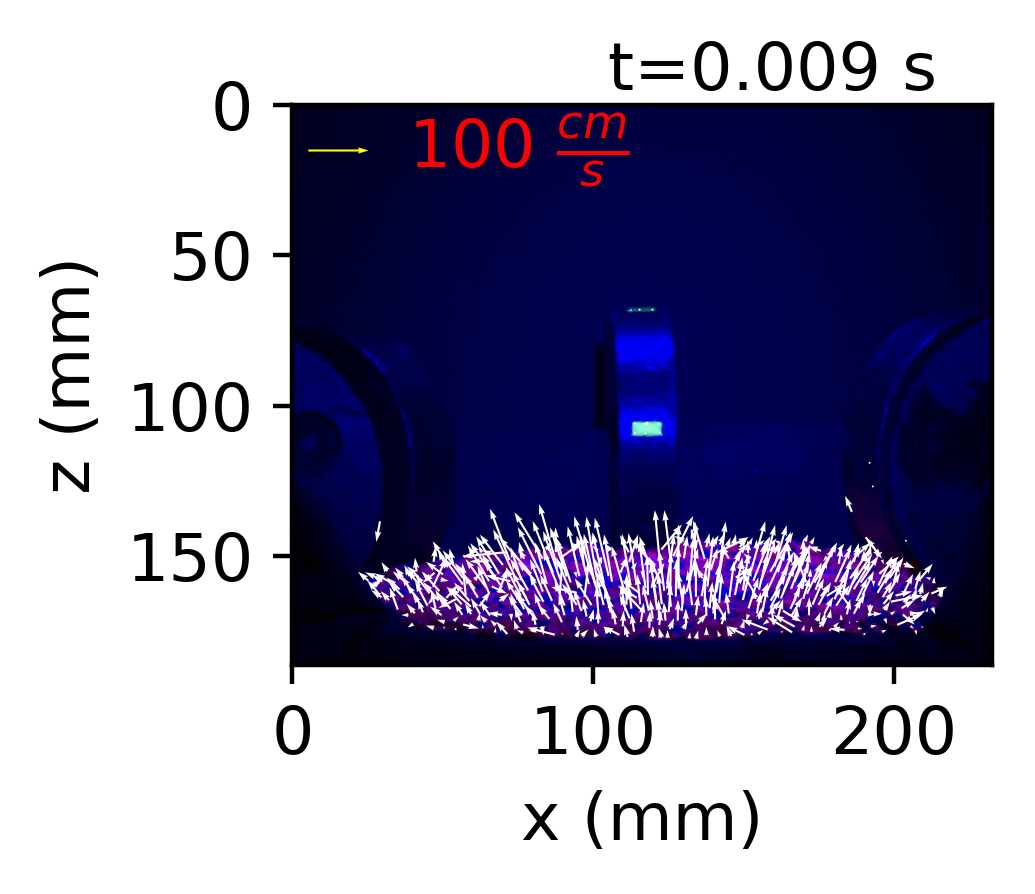} 
\centering\includegraphics[width=2.5in]{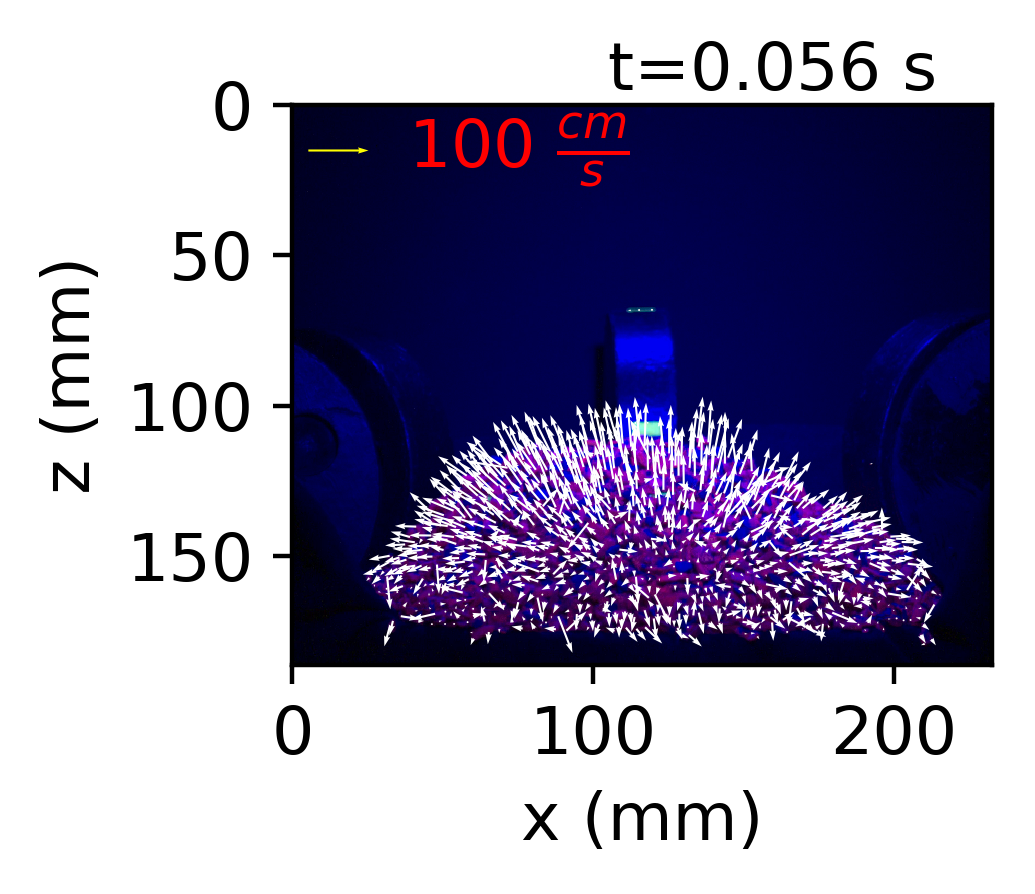} 
\centering\includegraphics[width=2.5in]{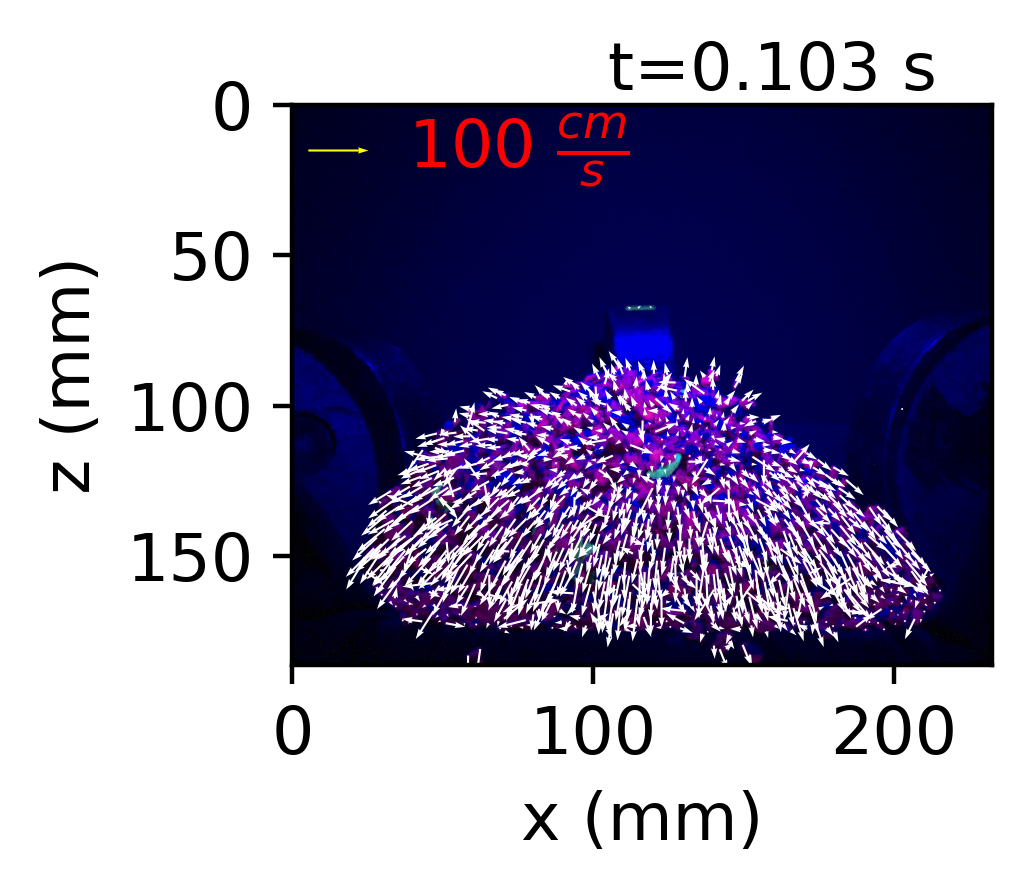} 
\centering\includegraphics[width=2.5in]{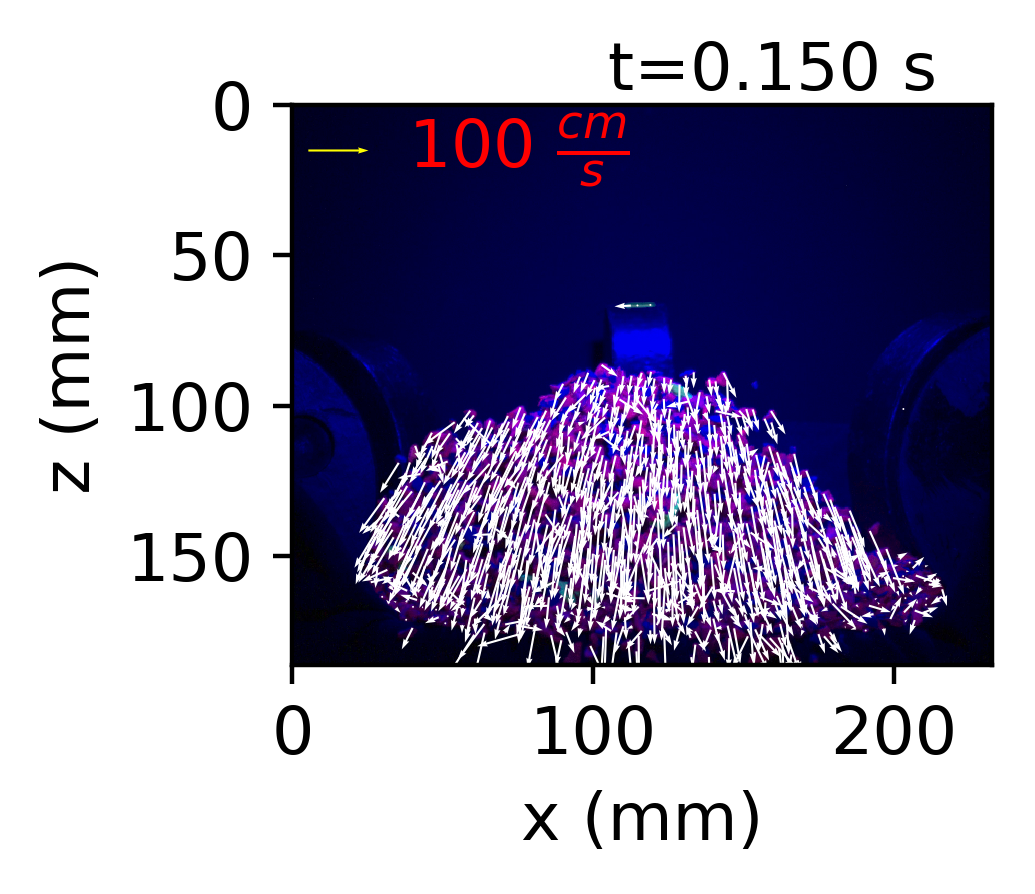} 
\centering\includegraphics[width=2.5in]{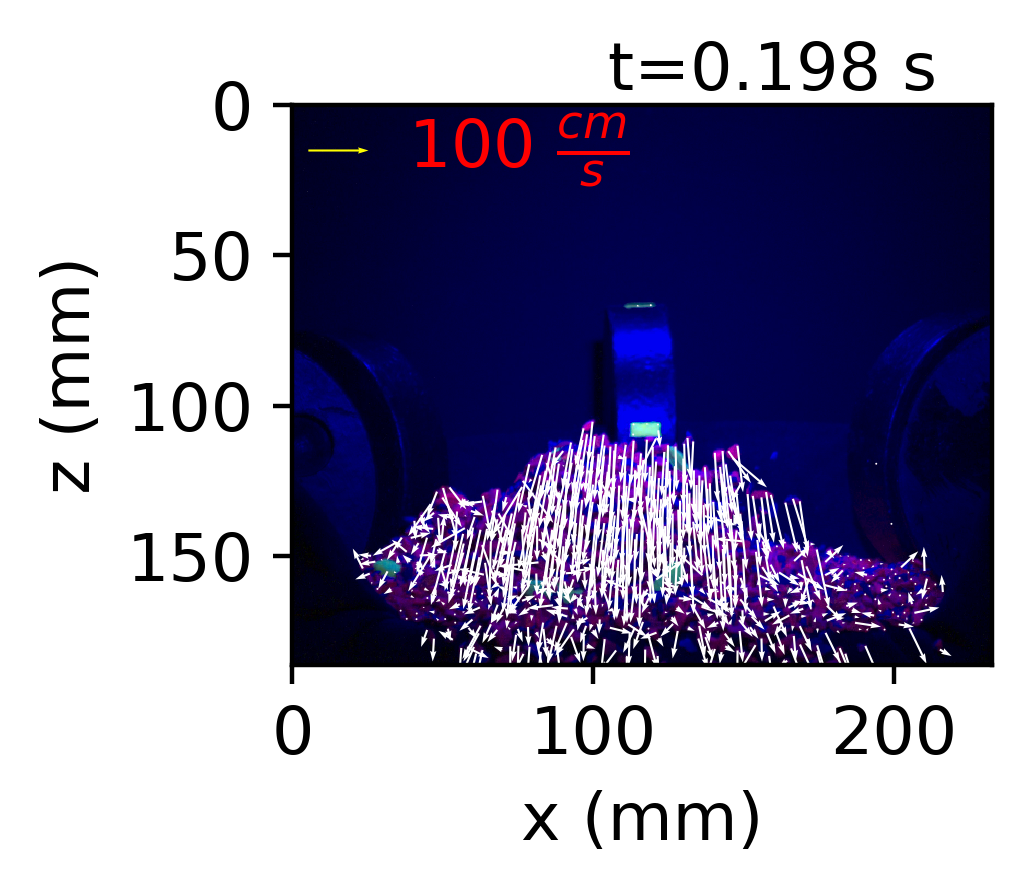} 
\centering\includegraphics[width=2.5in]{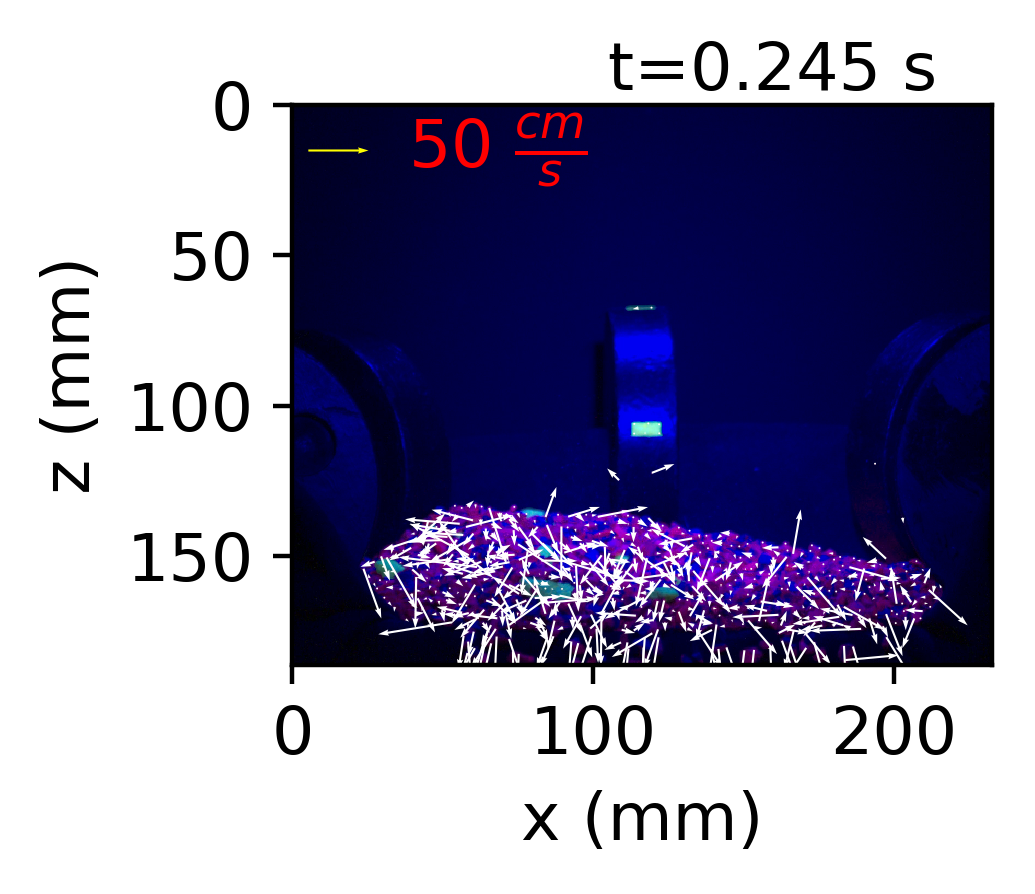} 
\caption{High speed video snapshots of experiment 42 and similar to Figure \ref{fig:hs39}.
}
\label{fig:hs42}
\end{figure*}

\begin{figure*}
%\centering\includegraphics[width=3in]{43_a.png}  % this block is for two column layout
%\centering\includegraphics[width=3in]{43_b.png} 
%\centering\includegraphics[width=3in]{43_c.png} 
%\centering\includegraphics[width=3in]{43_d.png} 
%\centering\includegraphics[width=3in]{43_e.png} 
%\centering\includegraphics[width=3in]{43_f.png} 
\centering\includegraphics[width=2.5in]{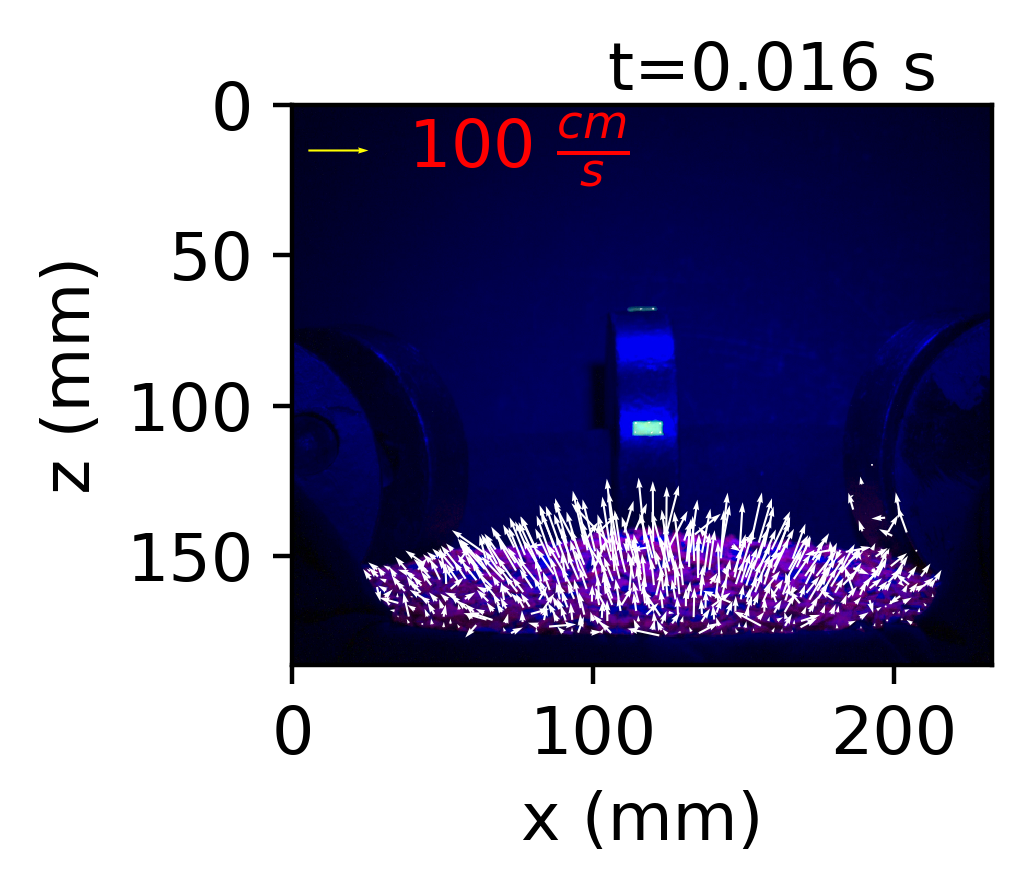} 
\centering\includegraphics[width=2.5in]{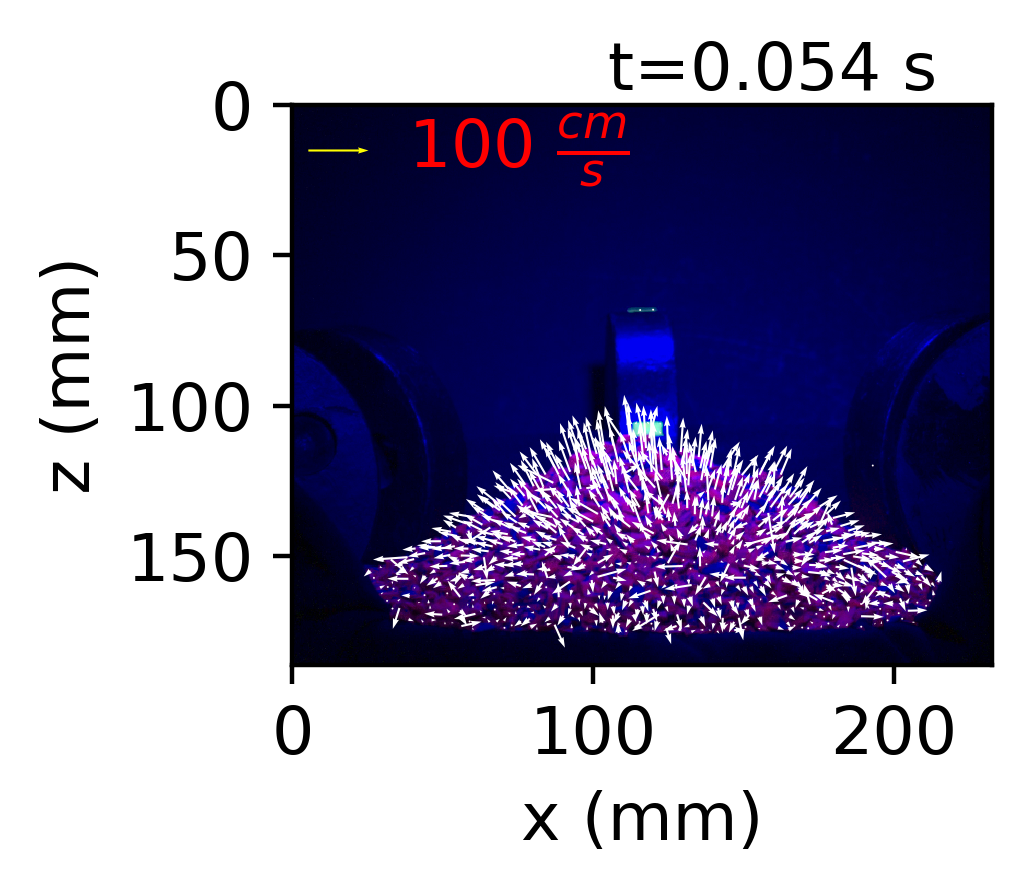} 
\centering\includegraphics[width=2.5in]{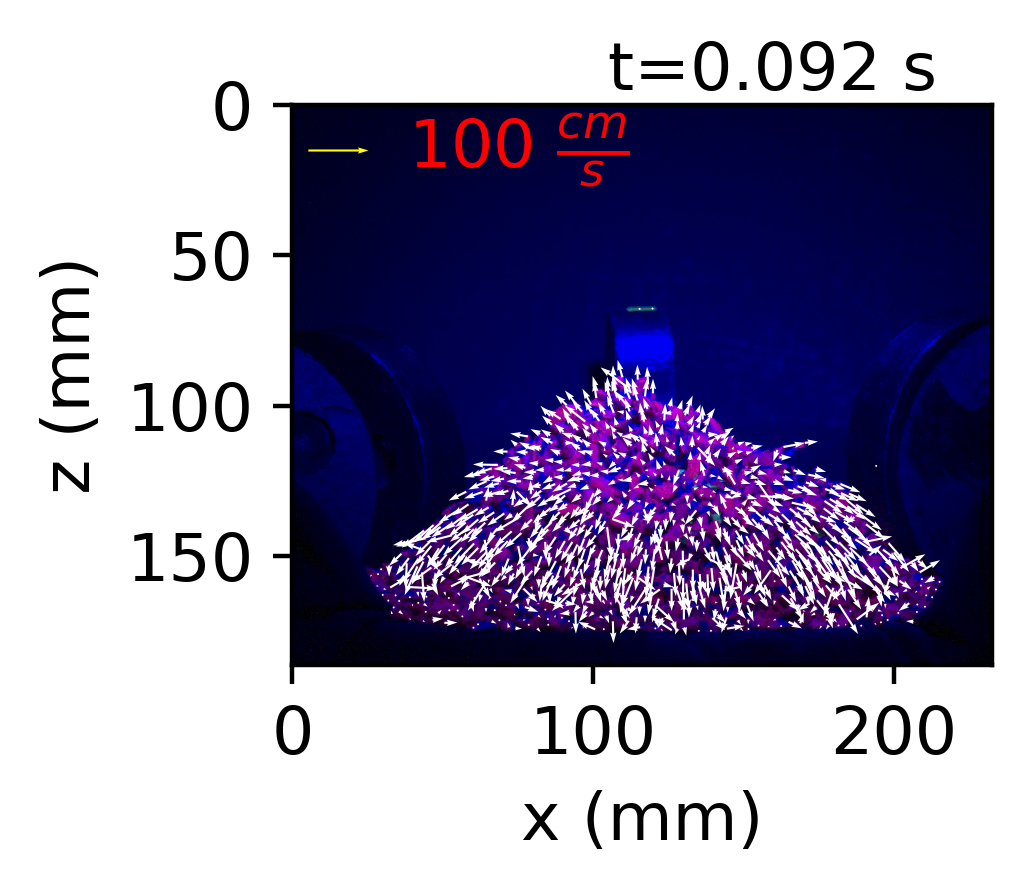} 
\centering\includegraphics[width=2.5in]{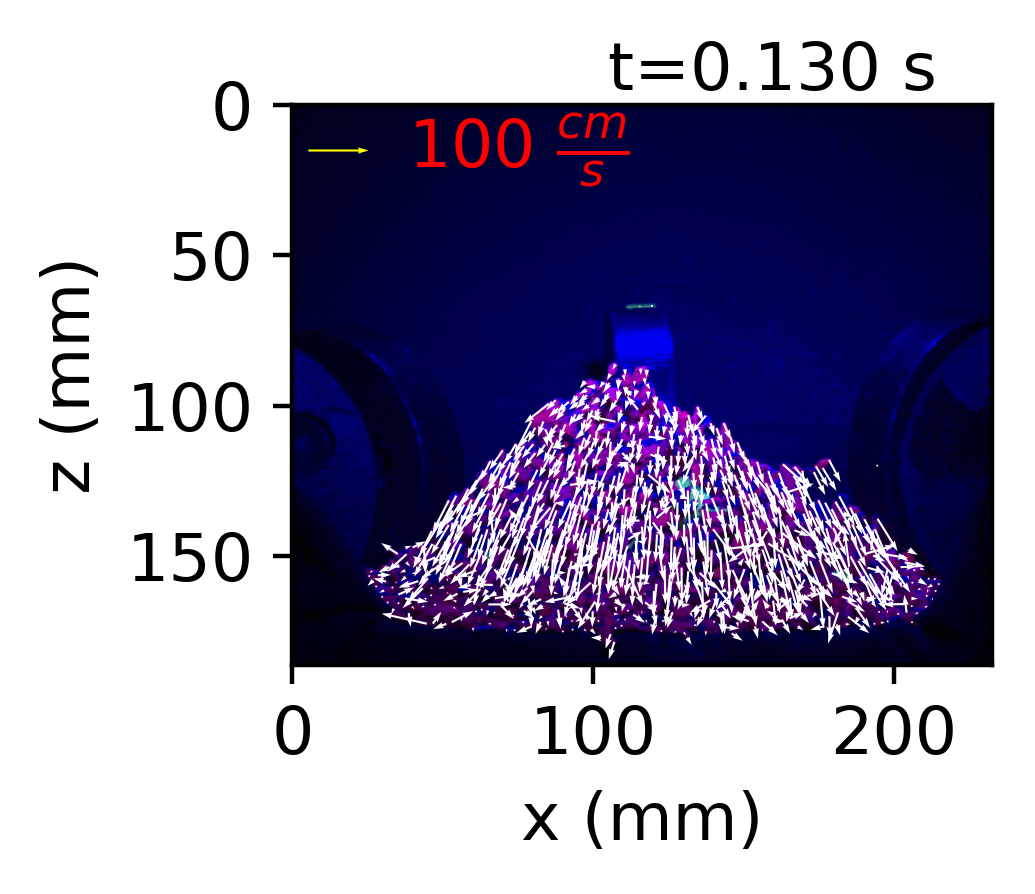} 
\centering\includegraphics[width=2.5in]{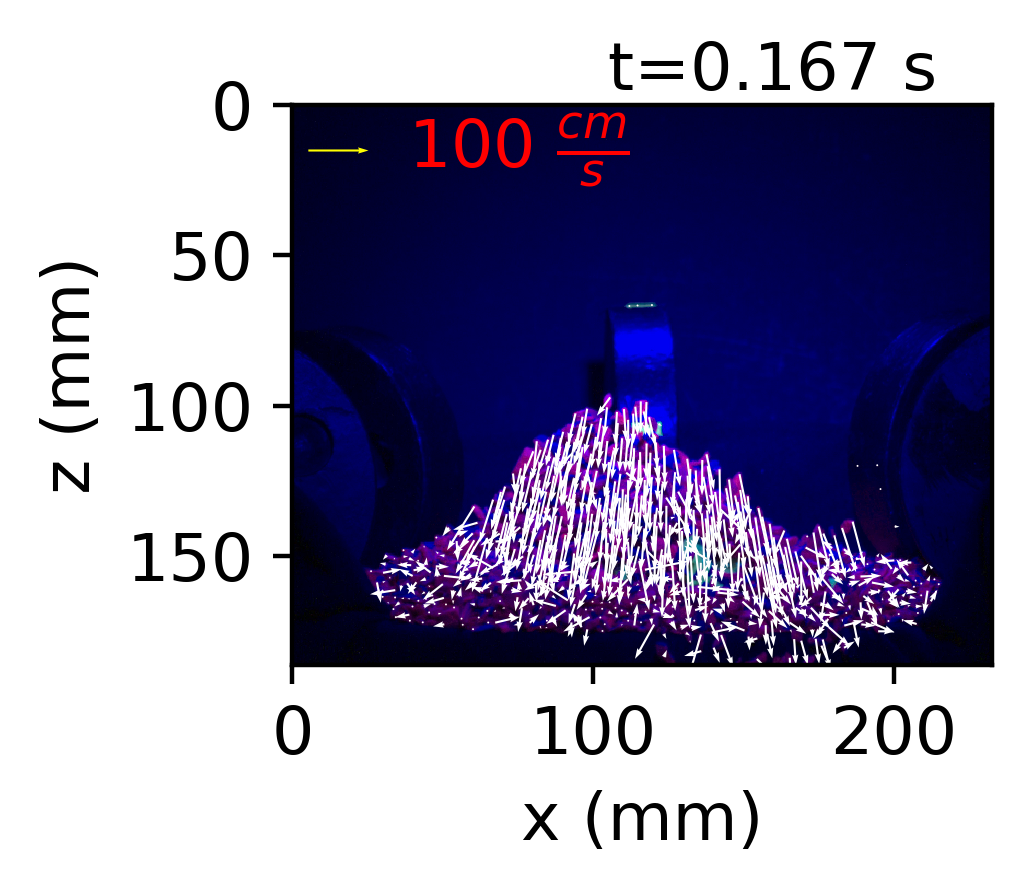} 
\centering\includegraphics[width=2.5in]{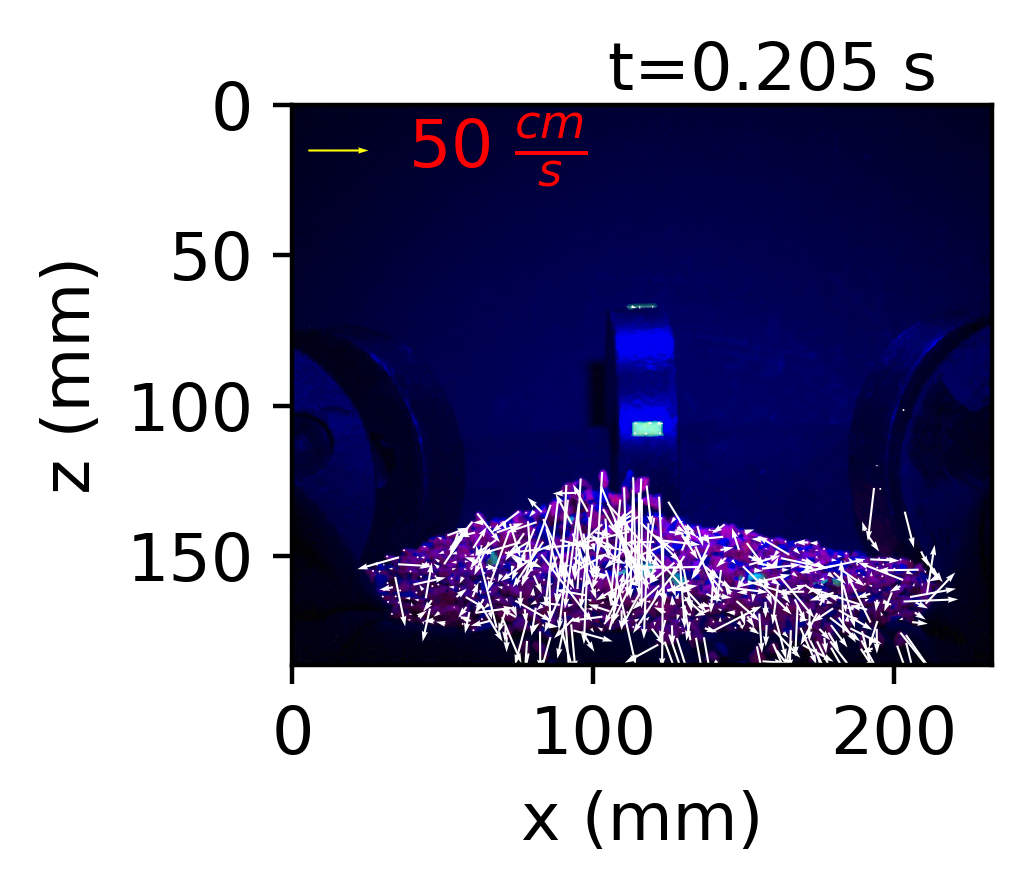} 
\caption{High speed video snapshots of experiment 43 and similar to Figure \ref{fig:hs39}.
}
\label{fig:hs43}
\end{figure*}

\begin{figure*}[tb] %star notation lets figure ignore column settings at top of document
    \centering
    \begin{subfigure}[t]{0.4\textwidth}
        \centering
        \includegraphics[width=\textwidth]{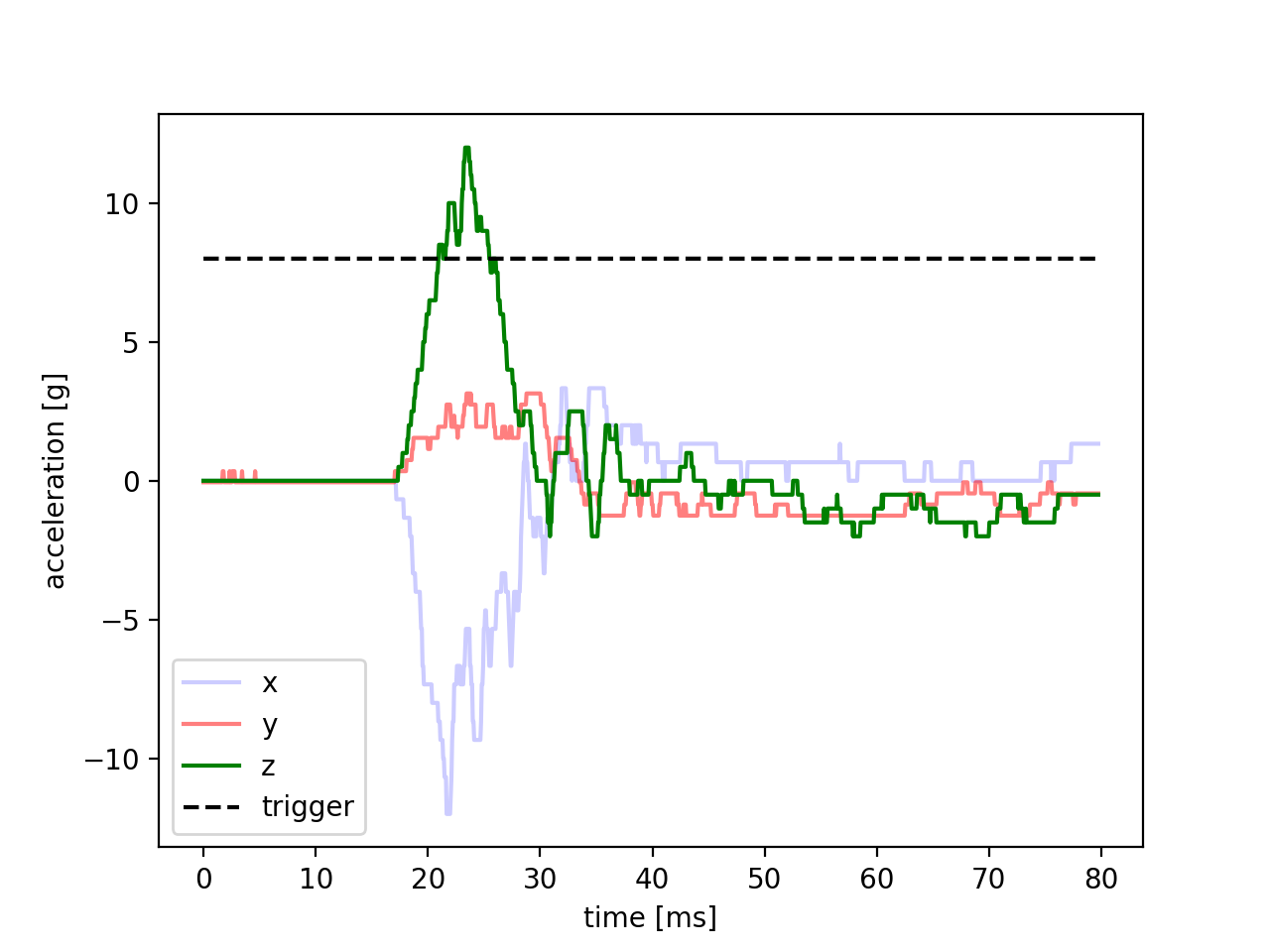}
        \caption{39}
        \label{fig:acc_39}
    \end{subfigure}
    \begin{subfigure}[t]{0.4\textwidth}
        \centering
        \includegraphics[width=\textwidth]{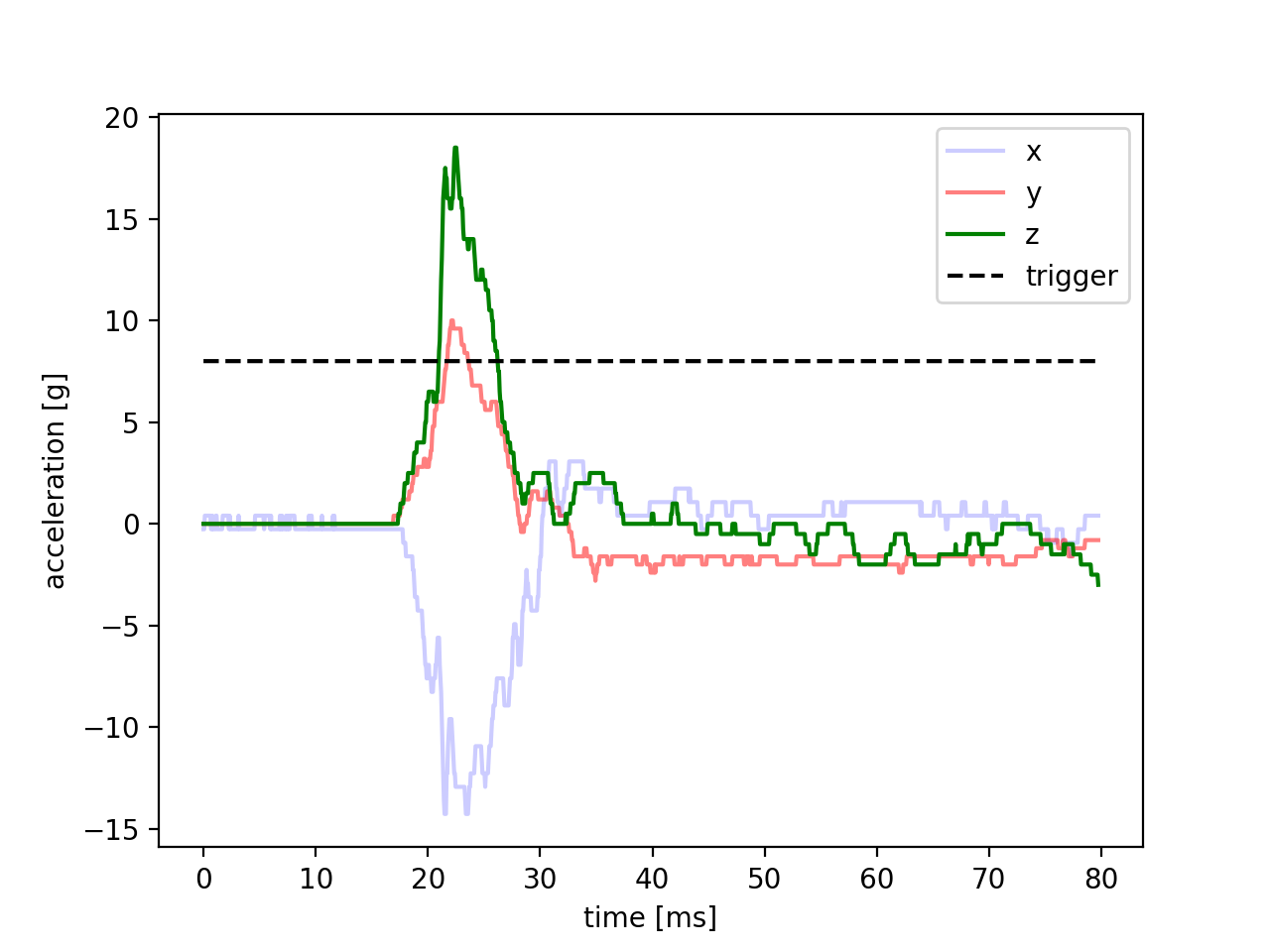}
        \caption{41}
        \label{fig:acc_41}
    \end{subfigure}
    \begin{subfigure}[t]{0.4\textwidth}
        \centering
       \includegraphics[width=\textwidth]{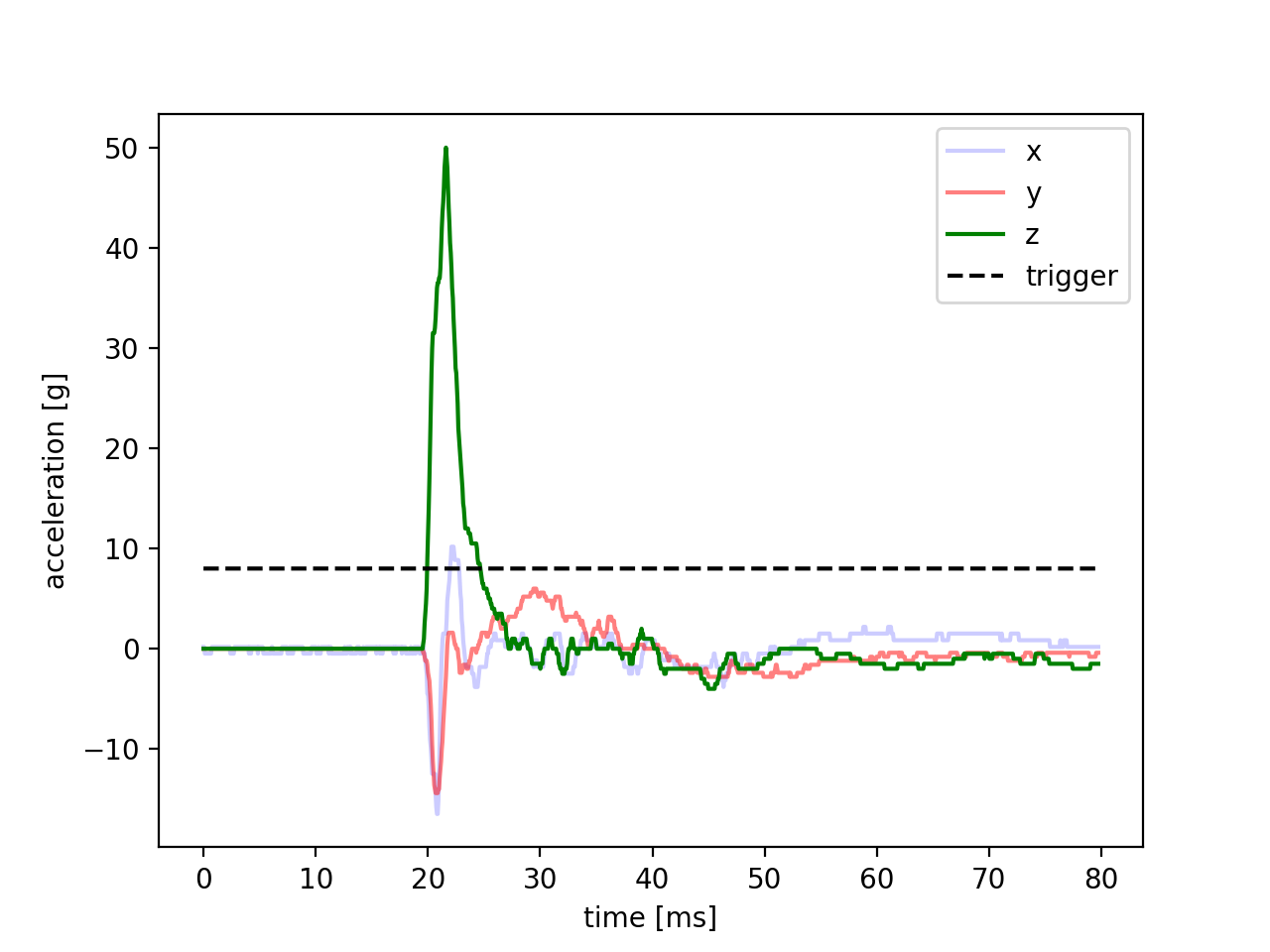}
        \caption{42}
        \label{fig:acc_42}
    \end{subfigure}
    \begin{subfigure}[t]{0.4\textwidth}
        \centering
        \includegraphics[width=\textwidth]{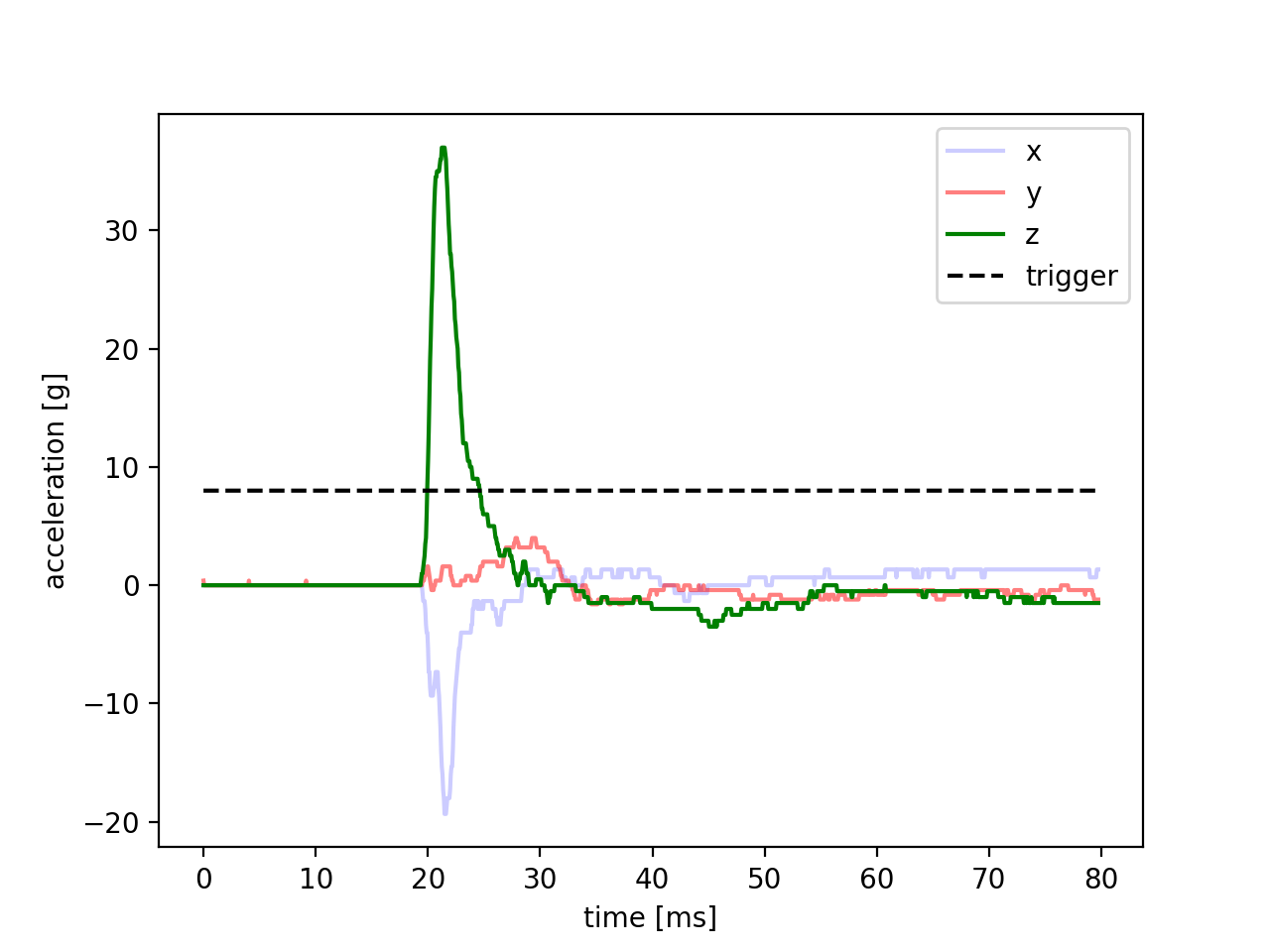}
        \caption{43}
        \label{fig:acc_43}
    \end{subfigure}
    \caption{Accelerometer measurements of subsurface pulse. Data was measured using an ADXL377 200g accelerometer buried 6 cm below the surface of the granular mixture and connected to an Arduino Mega.  Weaker pulses (> 20g) are shown in \subref{fig:acc_39}) and \subref{fig:acc_41}) and stronger pulses (< 30g) are shown in \subref{fig:acc_42}) and \subref{fig:acc_43}).  The accelerometer was positioned in the bowl so that the z-axis is in the vertical direction and the x and y axes are in a plane parallel to the surface of the granular mixture. %Pulse shapes were generated with a lever arm shown in Fig. \ref{fig:arm}.
    }
    \label{fig:accel}
\end{figure*}

While they are in the air, larger green particles have similar velocities to smaller red particles that in are proximity.  
Figure \ref{fig:std_ang} shows the standard deviation of the difference in the angles of the velocity vectors of the green and red particles as a function of the distance between one green particle and a red particle. The angles of the red particles' velocity vectors are subtracted from the angle of one green particle in the ejecta curtain. This angular difference is found for all red particles within some distance from the green particle. The standard deviation of the angular difference is then found and plotted as a function of this distance. The blue circles show the standard deviation calculated for the ejecta at a time when the curtain reached near max height and the orange diamonds at a time when the material is landing. 
Both times used the same green particle. For earlier times, when the ejecta is moving upward, the standard deviation of the difference in velocity directions is small for nearby particles and increases as more distant particles are included. Later in time the standard deviation is large for nearby particles and varies with increasing distance. This shows that red particles in close proximity to the green particle are moving together with similar velocities early in time and are scattering later in time.
 
We find that the ejecta velocities and trajectories are independent of particle size, as is true for crater ejecta \citep{hirata06}. 
Thus the tendency for large particles to be on top after particles land must be due to scattering
that happens upon landing.
We infer that the ballistic sorting process \citep{shinbrot17} is the reason that larger particles remain uncovered after all the ejecta has landed.

The fastest ejected particles are in the center of the bowl and directly above the impact site.
Particles that are more distant from the center land earlier than particles launched from the center of the bowl.
We suspect that particles originally below the surface are ejected with somewhat lower velocities than particles that are originally on or near the surface, as observed in the simulations by \citet{tancredi12}, though it is
difficult to track these lower particles as they are obscured by particles that leave the surface earlier.
The particles underneath that are launched later into the air, land earlier. 
Particles launched into the air earlier scatter of the lower particles as the ejecta lands. 
We do not see faster particles overtaking and scattering with slower particles in the air, suggesting that the magnitude of the velocities are similar.

To check that the pulse acceleration profiles are consistent with the velocity of ejected particles, we estimate the velocity of the pulse by taking the max acceleration and multiplying by the pulse duration. 
In this way, we calculate ejecta velocities of about 1.2, 1.2, 1.5, and 1.1 m/s for experiments 39, 41, 42, and 43 respectively.
Velocities measured in our experiments were lower by a factor of two from what was estimated. 
%which is expected. 
When the profile is integrated over the total time of the experiment, the ejecta velocities are approximately consistent with the velocities estimated from the acceleration profiles. 
This implies that the acceleration profiles
are consistent with the observed particle trajectories.
%The time for sound waves to travel across the depth of the bowl is less than the duration of our pulses.

The ejecta pattern due to a subsurface seismic pulse differs remarkably from that of an impact crater. 
An impact crater ejecta during a snapshot is in the shape of a cone because there is a relation between ejecta time, velocity and angle. 
This means that few particles are in proximity when they land. 
For a seismic pulse, nearby particles are launched a similar times and with similar velocities and so land in proximity. 
Thus we expect more particle-particle collisions in ejecta caused by seismic pulses than in crater ejecta.
Crater ejecta would scatter with particles already present on the surface, but would be less likely to scatter with each other.
A boulder present on the surface could remain there, but only a boulder that is launched in the
crater ejecta could become stranded on the surface.  In contrast, ejecta launched by a seismic pulse and containing boulders 
would suffer collisions between ejecta particles as well as surface particles upon landing.  

We estimate the coefficient of restitution from the ratio of particle's speed after they 
scatter (on landing)  to their descent speed just prior to scattering.
The approximate ratios of the vectors lengths  in Figures \ref{fig:hs39}-\ref{fig:hs42} (taking the ratio
of the vector lengths for a few particles) is about 1/3.
The coefficient of restitution is relevant for momentum exchange during particle-particle collisions.
Every collision reduces the relative normal speed by the restitution coefficient. 
If the coefficient of restitution is very low then small particles would stop moving after scattering
and they would only leave a larger particle uncovered if they were perched unstably on it after landing.
With a larger coefficient of restitution a small particle  can  be scattered far enough away  to leave a  larger particle
uncovered, though the distance travelled should exceed the width of the larger particle.

The pulse, as it travels through the medium, loses energy and spreads out due to particle-particle contacts and friction.
Ejection velocities are  greatest at the center of the ejecta plume and decrease further away from the center. 
The initial velocity vectors are not vertical away from the center.  
This means that particles are on diverging trajectories and particles tend to move or be scattered away from the center, making it possible to strand a larger particle in the center of the bowl than near the rim. 
This implies that the shape of the seismic waves in our experiments are spherical.
Pulse propagation through an asteroid may be inconsistent due to interior properties changing the pulse shape. 
However, so long as the pulse is strong enough to accelerate material off of the surface then some of the phenomena seen here might also arise on them.

We also found that once a large particle is stranded it tends to stay there and not be reburied. 
The number of large particles stranded on the surface was dependent on its depth. 
A single pulse can excavate a boulder if the acceleration of the seismic pulse is greater than the surface gravity (no material is ejected), vertical distance of the ejected material is greater than the boulder height (ejected material is above the large boulder and thus can scatter on landing), and the depth of material ejected is larger than the boulder height (otherwise boulder stays buried). 
We use these three conditions in Section \ref{sec:asteroid} to predict a regime for boulder stranding. 

When multiple pulses, separated in time, are sent through the gravel mix we found that they continue to excavate large particles. Once the particles are on the surface they tend to remain there after multiple pulses. Due to the absence of material above the large particle in the ejecta curtain nothing is present to bury the boulder upon landing. 
%However, if there is a large amount of smaller particles above the large particle in the ejecta curtain then the large particles can be reburied as the smaller particles do not scatter sufficiently far to keep the boulder uncovered.
If the acceleration of the pulse is below the surface gravity then no material is lofted into the air, but size sorting is still observed. In this case, the BNE is driven by a ratcheting effect
%as discussed by XXXXCITATIONXXXX
versus a ballistically sorted BNE.

In summary, our laboratory experiments illustrate that a single subsurface pressure pulse that lofts surface material can leave previously buried large particles on the surface. The mechanism is robust as every experiment we ran showed the phenomenon. 
Particle trajectories show that nearby particles have similar velocity vectors while particles are in the air but are randomly oriented upon landing when they scatter.
Nearby but different sized particles also exhibited similar trajectories while in the air, implying that larger particles were left on the surface because of collisions that took place upon landing.  

% SD of difference in angles as a function of seperation distance

% Impact 39 figure

% Impact 41 figure

% Impact 42 figure

% Impact 43 figure

% Accelerometer images

\begin{table*}
\centering
\caption{\large  Nomenclature \label{tab:nomen}}
\begin{tabular}{@{}lllllll}
\hline
%Seismic amplification factor & $S$ \\
Mass of asteroid & $M$ \\
Radius of asteroid & $R$ \\
Density of asteroid & $\rho$ \\
%Density of projectile & $\rho_{\rm proj} $ \\
Surface gravitational velocity & $V_{\rm grav} = \sqrt{GM/R}$ \\
Surface gravitational acceleration & $g = GM/R^2$ \\
Diameter of asteroid & $ D$ \\
\hline
Diameter of projectile & $D_{\rm proj}$ \\
Mass of projectile & $m_{\rm proj} $\\
Velocity of projectile & $V_{\rm proj} $\\
%Force impulse function & $F_s(t)$ \\
Kinetic energy of projectile & $E_{\rm proj} = \frac{1}{2} m_{\rm proj} V_{\rm proj}^2$ \\
Total radiated seismic energy &  $E_s$ \\
Seismic efficiency &$ \epsilon_s = E_s/E_{\rm proj}$ \\
\hline
Diameter of crater & $ D_{\rm crater}$ \\
P-wave speed & $V_p$ \\
Seismic source duration & $\tau_s$ \\
Peak seismic frequency & $f_s \sim 1/\tau_s$ \\
Exponent for impact scaling & $\mu $ \\
Depth of ejected material & $h_{\rm ejected}$ \\
Maximum height reached by ejecta  & $h_{\rm max}$ \\
Boulder height & $h_{\rm boulder}$\\
Distance from impact site & $d$ \\
Travel attenuation function & $f_{\rm travel}(d)$ \\
Amplitude of velocity displacement in pulse   & $v_u(d)$ \\
Amplitude of velocity displacement in pulse near crater & $A_u$ \\
Source time parameter & $X_s$ \\
\hline
\end{tabular}
\end{table*}

\section{Regime for Boulder stranding via an impact excited seismic pulse}
\label{sec:asteroid}

In the previous section we presented laboratory experiments showing that larger particles that are previously buried can emerge on the surface following a single upward propagating subsurface pressure pulse that lofts particles into the air.
In this section we consider the possibility that a seismic pulse excited from an energetic impact can cause boulders to be stranded on rubble pile asteroids such as 162173 Ryugu, 101955 Bennu and 25143 Itokawa. We assume that both the asteroid and impactor are spherical and the collision is head-on.
Nomenclature used in this section is listed in Table \ref{tab:nomen}.

We describe the excitation of a seismic pulse from a meteoroid impact in terms of two parameters, 
the seismic efficiency, $\epsilon_s$, and the seismic source time $\tau_s$.
As these parameters are poorly constrained,  instead of assuming approximate values for them, we search for a regime
that allows boulders to be stranded on the surface.
%Using scaling relations by \citet{housen11} in the gravity regime, appropriate for a rubble pile,  we estimate the crater diameter.  
We estimate the amplitude $A_u$ of the velocity of 
displacement motions in the seismic pulse.   To correct for travel distance and spreading of energy , we assume that the pulse amplitude 
\begin{equation}
v_u(d) = A_u f_{\rm travel} \left(\frac{D_{\rm crater}}{d} \right) \label{eqn:vud}
\end{equation}
depends on a function $f_{\rm travel}$ of the distance traveled $d$, but we neglect
dispersion and so do not vary the pulse duration $\tau_s$.
Simulations by
\citet{tancredi12} find that particles ejected from the surface by a pressure pulse have ejection velocities
similar to $v_{\rm eject} \sim v_u$, the displacement velocity amplitude in the pulse.  We estimate the depth of lofted
material $h_{\rm eject} $ from the ejection velocity and the kinetic energy per unit area in the seismic pulse.   
Our experiments suggest that 
a boulder can be stranded on the surface if 
\begin{itemize}
\item
Accelerations in seismic pulse excited by the impact are above that of surface gravity.
\item 
The vertical distance traveled $h_{\rm max}$ by
material lofted above the surface is greater than the boulder height, $h_{\rm boulder}$.
\item
The depth of material lofted $h_{\rm eject}$ is larger than the boulder height.
\end{itemize}

The kinetic energy of the projectile $E_{\rm proj} = \frac{1}{2} m_{\rm proj} V_{\rm proj}^2$ 
with mass $m_{\rm proj}$ and velocity $V_{\rm proj}$ 
can be compared to the total radiated seismic energy, $E_s$, giving a
seismic efficiency factor 
\begin{equation}
\epsilon_s \equiv \frac{E_s}{E_{\rm proj}}. 
\end{equation}
%The seismic efficiency $\epsilon_s$ for impacts onto asteroids is not well known. 
Estimates for the seismic efficiency
range from $\epsilon_s \sim 10^{-2}$ to $10^{-6}$ 
(see experiments, simulations and discussions by 
\citealt{mcgarr69,schultz75,melosh89,richardson05,shishkin07,lognonne09,yasui15,guldemeister17}).

For a pressure pulse with
half width  (in time) $\tau_s$, the width in space of the traveling seismic pulse is $ \sim V_p \tau_s$
where $V_p$ is the p-wave speed.
The total energy per unit volume in the seismic pulse emitted at the crater base is proportional to $\rho A^2_{u}$.
%The area of the crater is $\pi D^2_{\rm crater}/4$.
The volume of displaced material is given by the area of the crater, $\pi D^2_{\rm crater}/4$, times the spatial width of the pulse. The seismic energy is therefore given by
\begin{equation}
E_s \sim \rho A_u^2 \frac{\pi D_{\rm crater}^2}{4} V_p \tau_s.
\end{equation}
Using the seismic efficiency and projectile mass and velocity,
the amplitude in displacement velocity of the seismic pulse as it leaves the crater base
\begin{equation}
A_u^2  \sim V_{\rm proj}^2  \frac{\epsilon_s}{3} \frac{D_{\rm proj}}{V_p \tau_s}   \left( \frac{D_{\rm proj}}{D_{\rm crater}} \right)^2 \label{eqn:Au1}
\end{equation}
and we have assumed similar densities for asteroid and projectile;  $\rho_{\rm proj} \sim \rho$.
A particle ejected from the surface at velocity $v_{\rm eject} = v_u$ at distance $d$ from the impact site
would reach a maximum height of travel 
\begin{equation}
h_{\rm max} (d) \sim \frac{v_u^2}{2g}  = \frac{R A_u^2}{2 V_{\rm grav}^2} f_{\rm travel}^2(d) \label{eqn:hmax1}
\end{equation}
where $g$ is the gravitational acceleration at the surface. 

In an elastic regime the function describing the decay of the seismic pulse amplitude
$f_{\rm travel} (d) \propto d^{-1}$ depends on the distance propagated. 
Scaling from the crater diameter and setting $d = D$, the asteroid diameter, for
a pulse reaching the impact site's antipode, we assume 
$f_{\rm travel} (D) \sim D_{\rm crater}/D$.  Inserting this into equation \ref{eqn:hmax1}  and using equation 
\ref{eqn:Au1}, %for the crater diameter 
we estimate the ejecta reaches a vertical height above initial position

\begin{equation}
\left( \frac{h_{\rm max}}{R} \right) |_{d=D} = \frac{\epsilon_s}{6} \left( \frac{D_{\rm proj}}{V_p \tau_s} \right) \left( \frac{V_{\rm proj}} {V_{\rm grav}} \right)^2
\left( \frac{D_{\rm proj}} {D} \right)^2 .
\end{equation}

We can invert this equation giving a projectile diameter 
\begin{equation}
\left( D_{\rm proj,hmax} \right) |_{d=D}  \sim  D  \left( \frac{6}{\epsilon_s } \frac{h_{\rm max} }{R}  \frac{V_p \tau_s} {D_{\rm proj}}\right)^\frac{1}{2}
\left( \frac{ V_{\rm grav}}{V_{\rm proj}} \right) \label{eqn:D_hmax3}
%&=  D  \left( \frac{3}{4\epsilon_s }
%\frac{h_{\rm max} }{R} \frac{h_{\rm ejected} }{R} \right)^\frac{1}{3}
%\left( \frac{ V_{\rm grav}}{V_{\rm proj}} \right)^\frac{2}{3}. 
\end{equation}
in terms of the ejecta height.
%Equation \ref{eqn:D_hmax2} gives an estimate for a projectile that can strand a boulder near the impact site with $f_{att}(d) \sim 1$, and
Equation \ref{eqn:D_hmax3} gives an estimate for a projectile that could strand a boulder on the impact site's antipode. 

The velocity at which material is ejected from the surface  should be similar to the displacement velocity of the seismic pulse
\begin{equation}
v_{\rm eject} \sim v_u
\end{equation}
%(see section 5 by \citealt{tancredi12}).
\cite{tancredi12} in section 5 simulated the maximum height of ejecta from a sub-surface pulse. Using equation \ref{eqn:hmax1} and the velocity of our pulses we calculate max heights of 7, 7.5, 11, and 6 cm for experiments 39, 41, 42, and 43 respectively. These max heights have a similar order as given in Figure 15 of \citealt{tancredi12} supporting that the ejection velocity is similar to the displacement velocity of the pulse.

The  kinetic energy per unit area of ejected material is given by
\begin{equation}
e_{\rm ejected} \sim  \frac{1}{2} h_{\rm ejected} \rho v_u^2.
\end{equation}
The kinetic energy per unit area in the seismic pulse can be estimated by integrating over the pulse at a moment before it reaches the surface,
$e_{\rm pulse}\sim  \int dx \  \rho v_u^2$.  
Assuming that the pulse travels with velocity $V_p$ and that it's width has not spread in time (i.e. the pulse maintains it's shape),
\begin{equation}
e_{\rm pulse} \sim  \rho v_u^2 V_p \tau_s.
\end{equation}
Equating these two estimates for kinetic energy gives us an estimate for the depth of ejected material
\begin{equation}
h_{\rm ejected} \sim 2V_p \tau_s. \label{eqn:hej}
\end{equation}

The contact-and-compression 
phase of an impact excites a hemispherical shock wave in the ground that propagates 
away from the impact site \citep{melosh89}. 
As the shock wave propagates, it  degrades into a purely elastic (seismic) wave. 
The structure of the elastic wave is expected to be complex, with multiple pulses
associated with the elastic precursor to the shock wave, an elastic remnant to a plastic wave during
the transition between shock and elastic wave, and reverberations associated with different
seismic impedances in the target, rock fractures and compactification  (see section 5.2.6 by \citealt{melosh89}).
Low velocity laboratory impacts into granular media  measure source times (as a half width) 
of about 10 $\mu$s \citep{yasui15}.  Similar durations are measured in sandstone targets \citep{hoerth14} and are predicted via
numerical simulation \citep{guldemeister17}.  
These pulse durations may be shorter than  excited by astronomical impacts, as the seismic source time $\tau_s$ could be longer for more energetic impacts (e.g., \citealt{lognonne09}).
Missile impacts estimate peak seismic frequencies in the range $f_s \sim $ 10--40 Hz \citep{latham70} and hydrodynamics 
simulations of asteroid impacts estimate a similar range   \citep{richardson05}.  
We relate source time to frequency with $\tau_s \sim 1/f_s$.  

To help identify the regime that allows boulders to be stranded we tentatively 
adopt a linear scaling between projectile radius and seismic source  time, defining a parameter 
\begin{equation}
X_s \equiv \frac{V_p \tau_s}{D_{\rm proj}}
\end{equation}
similar to equation 5 by \citet{lognonne09} that is based upon calculations of seismic 
power radiated into a homogeneous elastic half-space
\citep{wolf44,mcgarr69}.  Setting seismic source time parameter $X_s$ 
and requiring that $h_{\rm boulder} > h_{\rm ejected}$, equation \ref{eqn:hej} 
limits the projectile radius
\begin{equation}
D_{\rm proj,heject} \gtrsim  \frac{D}{4X_s} \frac{h_{\rm boulder}}{R}. \label{eqn:DE}
\end{equation}
With $h_{\rm max}>h_{\rm boulder} $ and our definition for source time parameter $X_s$, we rewrite
 equation \ref{eqn:D_hmax3} 
\begin{equation}
\left( D_{\rm proj,hmax} \right) |_{d=D}  \gtrsim  D  \left( \frac{6}{\epsilon_s } \frac{h_{\rm boulder} }{R}  \right)^\frac{1}{2} X_s^{\frac{1}{2}}
\left( \frac{ V_{\rm grav}}{V_{\rm proj}} \right),\label{eqn:DM}
%&=  D  \left( \frac{3}{4\epsilon_s }
%\frac{h_{\rm max} }{R} \frac{h_{\rm ejected} }{R} \right)^\frac{1}{3}
%\left( \frac{ V_{\rm grav}}{V_{\rm proj}} \right)^\frac{2}{3}. 
\end{equation}
giving an estimate for the size of projectile that could strand a boulder on the impact site's antipode.
A comparison between equation \ref{eqn:DE} and \ref{eqn:DM} shows there is a balance involving the seismic source time.
If the source time is short, then little material is ejected, but ejected material reaches a larger height.
If the source time is long, then more surface material is lofted, but not very high above the surface. 

Equation \ref{eqn:DE} gives a projectile diameter capable of causing a boulder at depth $h_{\rm boulder}$  be ejected from the surface.
Equation \ref{eqn:DM} gives the diameter of a projectile capable of causing ejecta to reach a height of $h_{\rm boulder}$. 
For a boulder of height $h_{\rm boulder}$ to be stranded on the surface we assume 
that both conditions must be satisfied.

\begin{figure*}
\centering\includegraphics[width=5in]{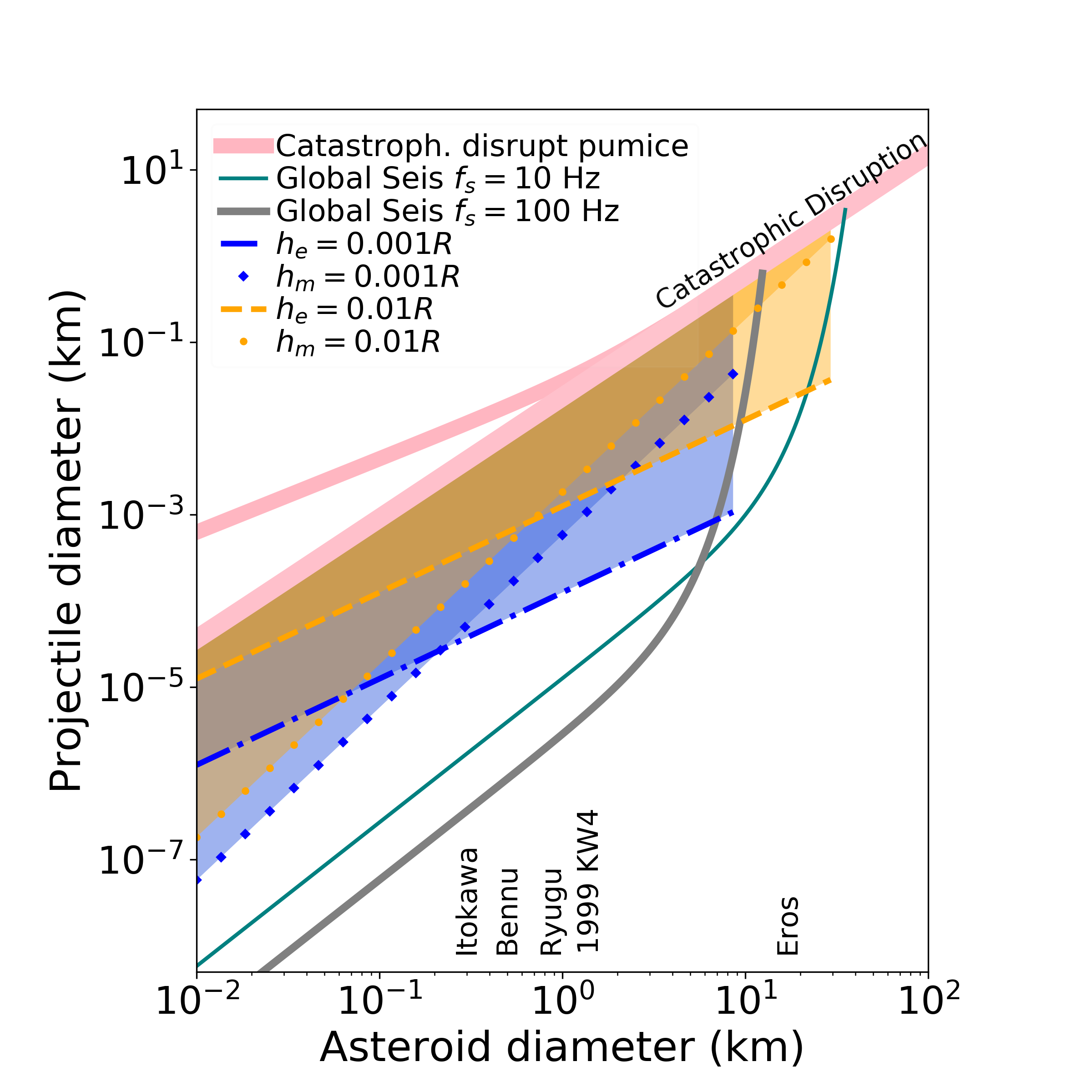} 
\caption{
Scaling for impacts on asteroids.  We show diameters of projectiles
capable of catastrophic disruption (wide pink lines), global seismic shaking (green and grey solid lines) and for stranding boulders
(orange and blue lines, dots and shaded regions).
Above the dashed orange line,  impacts cause seismic pulse ejected material to reach a height  $h_{\rm max} > 0.01 R$,
where $R$ is the radius of the asteroid.
Above the orange dotted line, the depth of ejecta  $h_{\rm ejected} > 0.01 R$.  We shade in orange the region above
these lines and the catastrophic disruption line.  The darker orange region is above 
 both dotted and dashed orange lines, and is where 
boulders with height $h_{\rm boulder} > 0.01R$ could be stranded on the surface by a single seismic pulse.
The dashed orange and dot dashed blue lines show equation \ref{eqn:DM} and the dotted lines are equation \ref{eqn:DE}, 
computed using seismic efficiency $\epsilon_s = 10^{-4}$
and with seismic source time  parameter $X_s = 2$.
Blue dot-dashed line,  line of diamonds and shaded areas show similar heights but for $0.001R$.
%Catastrophic disruption and global reverberation thresholds for impacts are also shown in this figure.
The wide pink lines shows the catastrophic disruption threshold as a function of asteroid diameter, computed
using Eq. (3) by \citet{jutzi10} and coefficients for pumice from their Table 3.  Two lines
are shown on the left, the upper line shows a curve for a body with strength and the lower curve for rubble. 
Two lower lines (thin green, and thicker grey) give minimum impactors capable of causing global seismic shaking (GS)
or seismic waves with accelerations
greater than the surface gravity. These are computed using equation 9 by \citet{richardson05} and are computed
for seismic frequencies 10 and 100 Hz. 
}
\label{fig:loglines}
\end{figure*}

In Figure \ref{fig:loglines} we plot 
the projectile diameter as a function of asteroid diameter that would strand a boulder on the surface from a seismic pulse
 that ejects surface particles.  Orange dashed and blue dot-dashed  lines show equation \ref{eqn:DM} computed with
 $h_{\rm boulder}/R = 0.01$ and 0.001, respectively.   Orange and blue dots  show equation \ref{eqn:DE}
 computed with the same ratios.
Above the orange dots, the depth of ejecta  $h_{\rm ejected} > 0.01 R$.  
Above the orange dashed line, the ejecta  reaches a height $h_{\rm max} > 0.01 R$.  
Above dotted and dashed orange lines, shaded in darker orange,
boulders with height $h_{\rm boulder} > 0.01 R$ could be stranded on the surface by a seismic pulse.
Above dotted and dot-dashed blue lines, shaded in darker blue, boulders with height $h_{\rm boulder} > 0.001R$ 
could be stranded on the surface by a seismic pulse.
We have assumed a  seismic efficiency $\epsilon_s = 10^{-4}$ (a relatively large value), density $\rho=1$ g/cm\textsuperscript{3}, projectile velocity $V_{\rm proj} = 5 $ km/s
(typical of asteroid collisions; \citealt{bottke94}), 
and a seismic source time parameter $X_s = 2$.

Catastrophic disruption and global reverberation thresholds for impacts are also shown in Figure \ref{fig:loglines}.
The wide pink lines shows the catastrophic disruption thresholds as a function of asteroid diameter, computed using Eq. (3) by \citet{jutzi10} and coefficients for pumice from their Table 3.  
Two lines are shown at the top. 
The upper one shows a body with strength and the lower curve extends the gravity regime for  a rubble asteroid.  
Assuming a single seismic wave frequency, \citet{richardson05} estimated  the diameter of a projectile $D_{\rm proj}$ (their equation 15) sufficient to cause seismic vibration across the whole body that is
above the surface gravitational acceleration. 
We show this global seismicity threshold as a thin green line for a seismic frequency of 10 Hz and as a thicker grey line for a frequency of 100 Hz.
%Taking seismic frequency $f_s \sim 1/\tau_s$, from the seismic source time and neglecting seismic attenuation,
%\citet{quillen19_bennu} similarly estimated a global seismicity threshold (their equation 21).  This is shown  as a dot-dash tan line on the same plot.  
For these curves, we assume  P-wave velocity $V_p = 100 $ m/s, typical of lunar regolith,
attenuation coefficient $Q = 2000$ and seismic diffusivity $K_s = 0.1\ {\rm km}^2 {\rm s}^{-1}$.

Figure \ref{fig:loglines} shows that larger impacts that are just below the catastrophic disruption threshold 
are capable of ejecting moderate depths of surface
material via excitation of a seismic pulse.    In  the darker orange region, boulders
larger than 1/100th the radius of the asteroid can emerge to the surface after the seismic
pulse ejects material off the surface. 
To eject surface material, the acceleration provided by the pulse once it reaches the surface must be greater than the net gravity at that point.
The shaded regions lie above the acceleration dependent global seismicity thresholds previously
estimated by \citet{richardson05} and so accelerations in the seismic pulse should satisfy this condition.

In Figure \ref{fig:loglines} we chose seismic source time with $X_s=2$.  Larger $X_s$ move the point where yellow dotted
and dashed lines cross upward and to the right on the plot.  The two lines cross because equation \ref{eqn:DE} has projectile
diameter inversely proportional 
to source time parameter $X_s$ whereas in equation \ref{eqn:DM} the diameter is $\propto X_s^\frac{1}{2}$.
For boulder stranding via impact excited seismic pulses to be a relevant process on asteroids, Figure \ref{fig:loglines} suggests
that the seismic source must satisfy $X_s \sim 1$.  
%This condition is equivalent to seismic pulse duration approximately equal to the sound travel time through the projectile. 
With  $X_s \sim 1$ then the time for the pulse to travel the diameter of the projectile is equal to the inverse pulse duration.
The time is longer than seen in some laboratory experiments \citep{yasui15} and numerical simulations \citep{guldemeister17} but shorter than computed via scaling estimates (used by  \citealt{lognonne09}).

Figure \ref{fig:loglines} shows requirements for boulder stranding at the antipode of an impact site.
Smaller impactors could strand the same size boulder nearer the impact side.  However we have
not taken into account attenuation of the seismic pulse as it travels through the asteroid but have allowed the energy of the pulse to spread out as a function of inverse distance. 
Pulse broadening and attenuation would increase the size of an impactor needed to strand a particular width of boulder on the surface.
Figure \ref{fig:loglines} shows that boulder stranding could only be accomplished by large and energetic impactors. 
A catastrophic impact that produced a cloud of debris could also leave large boulders on the surface as ejecta
re-accumulates.

\subsection{Ejecta Mass Fraction}
\label{sec:mass}

\begin{figure}
\centering\includegraphics[width=3.6in]{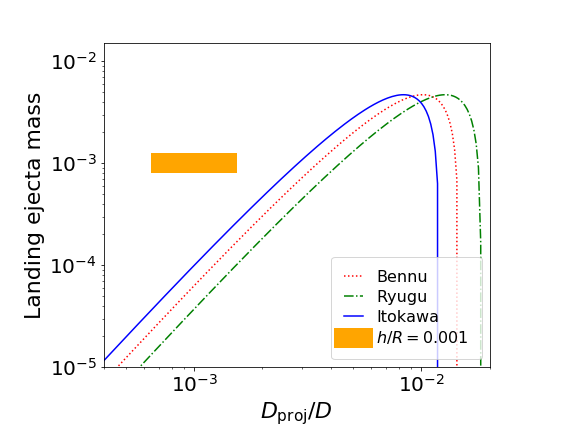} 
\caption{
The fraction of crater ejecta mass in units of asteroid mass as a function of asteroid to projectile diameter ratio
that is ejected below the escape velocity and  returns to
the asteroid surface.  This fraction is computed with equation \ref{eqn:fcrater} 
based on  scaling relations by \citet{housen11} and for three different diameter
asteroids.   The blue solid, red dotted and green dot-dashed lines show the fraction for asteroids with the diameter of Ryugu, Bennu and Itokawa, 
respectively.  The curves are computed with asteroid and projectile density $\rho=1$ g/cm\textsuperscript{3}, and projectile velocity $V_{\rm proj} = 5 $ km/s.
The orange bar shows a rough estimate for mass ejected due to a seismic pulse that originates from a strong but sub-catastrophic impact with 
mean ejecta depth $h/R = 0.001$.  This orange bar assumes that the impact lies in the blue and orange areas shown in Figure \ref{fig:loglines}.
If seismic pulses launched from impacts are strong enough to strand boulders on the surface, then
they could dominate over crater ejecta as a source of ballistically sorted material.
}
\label{fig:mass}
\end{figure}

In this subsection we compare the fraction of mass that falls onto an asteroid surface
as crater ejecta to that which could be lofted off the surface from a impact generated seismic pulse.
Both could be sources of material that is ballistically sorted upon landing.

Impact craters on a "rubble pile" asteroid may be in or near a gravity-scaling regime \citep{holsapple93,asphaug02}. For our order of magnitude analysis we assume an asteroid in the gravity-scaling regime.
Scaling relations for crater ejecta by \citet{housen11} (summarized in their Table 1 and with coefficients in their Table 3) describe mass in the ejecta as a function of position
(their equation 16), total mass ejected during crater formation (just above their equation 17),
and velocity of ejecta as a function of position (their equation 9 in the gravity regime).   
Together these give an estimate for 
 the total mass in ejecta  with velocities above the escape 
velocity $M(v> v_{\rm escape})$ during formation of a crater.   In units of the total ejected mass $M_{\rm crater}$, 
the fraction of crater ejecta mass that escapes the asteroid 
\begin{equation}
\frac{M( v> v_{\rm escape})}{M_{\rm crater}} \approx  
%\frac{1}{2} \left( \frac{8 \pi G \rho}{3}\right)^{- \frac{3\mu}{2}} \left( \frac{D_{\rm proj}}{D_{\rm crater}} \right)^3 \left( \frac{V_{\rm proj}}{R} \right)^{3\mu}, 
\left(\frac{D}{D_{\rm crater}} \right)^{-\frac{3\mu}{2}}
\end{equation}
with $M_{\rm crater} \sim 0.3 \rho R_{\rm crater}^3$.
The fraction of ejecta mass that falls back onto the surface in units of asteroid mass is
\begin{equation}
f_{\rm crater} \sim \frac{M_{\rm crater} - M( v> v_{\rm escape})}{M}  \nonumber \\
\sim 0.07 \left( \frac{D_{\rm crater}}{D}\right)^3 \left( 1- \left(\frac{D}{D_{\rm crater}} \right)^{-\frac{3\mu}{2}}\right). \label{eqn:fcrater}
\end{equation}
The diameter of the crater $D_{\rm crater} $ can be estimated from the projectile diameter using the scaling estimate for crater radius in
the gravity regime (also by \citealt{housen11})
\begin{equation}
\frac{D_{\rm crater} }{ D_{\rm proj}} \sim \left( \frac{ V^2_{\rm grav}} {V^2_{\rm proj}}  \frac{ D_{\rm proj}}{ D} \right)^{- \frac{\mu}{2 + \mu}} .
 \label{eqn:30}
\end{equation}

In Figure \ref{fig:mass} we show
the fraction of returning crater ejecta mass in units of asteroid mass computed with equation \ref{eqn:fcrater}
 as a function of asteroid to projectile diameter ratio.  The three curves give the fraction of mass
in crater ejecta   below the escape velocity.
The blue solid, red dotted and green dot-dashed lines show the fraction for asteroids with the diameter of Ryugu, Bennu and Itokawa, 
respectively.  
The curves are computed with asteroid and projectile density $\rho=1$ g/c\textsuperscript{3}, and projectile velocity $V_{\rm proj} = 5 $ km/s and exponent
$\mu = 0.4 $ following Table 3 by \citet{housen11}. 
Catastrophic impacts diameter ratio $D_{\rm proj}/D \sim 10^{-2}$ (see Figure \ref{fig:loglines}).
For reference, the regime of strong but subcatastrophic impacts with $D_{\rm proj}/D \sim 10^{-2}$
has a fraction of ejecta mass that falls back to the surface $f_{\rm crater} \sim 10^{-4}$.   

We now compare $f_{\rm crater}$ to the mass that is launched by a seismic pulse.
The mass fraction of material lofted via an impact excited seismic pulse is $\sim h/R$ where $h$ is the depth of material
ejected averaged over the surface.  We found in section \ref{sec:asteroid} that 
with seismic source time parameter $X_s \sim 1$, a strong but sub-catastrophic impact might launch a depth $h/R \sim 0.01 - 0.001$.
We place an orange bar on Figure \ref{fig:mass} for a mass fraction of $h/R = 0.001$ for a strong but sub-catastrophic impact with $D_{\rm proj}/D \sim 10^{-3}$.   
The orange bar lies above the mass fraction in crater ejecta.
If seismic pulses launched from impacts  are strong enough to uncover or strand boulders on the surface, then they could dominate over crater ejecta as a source of ballistically sorted material.

\section{Summary}
\label{sec:sum}

%\subsection{Discussion}

We have carried out laboratory experiments of impacts into polydisperse granular media, but focusing on impacts from below to mimic
the behavior of an impact excited seismic pulse reaching the surface of an asteroid.   
Our laboratory impacts are strong enough that particles
near and on the surface are ejected into the air.    The viscoelastic soft sphere simulations  by \citet{tancredi12} showed that
a strong pressure pulse excited by an impact would eject particles from an asteroid surface, and our experiments mimic this process.

We find that initially buried larger particles are often left on the surface after impact.
Using particle image velocimetry of  high speed video 
we measure ejecta velocities, finding that nearby particles have 
 similar velocities.   The time dependent ejecta velocity field is qualitatively different from crater ejecta curtains where there is a range
of ejecta velocities and strong correlations (and scaling relations) between
ejecta velocity, launch position and time (e.g., \citealt{housen11}).    
We find that our ejecta trajectories are independent of particle size until they land.
Collisions primarily take place upon landing and small particles scatter off of larger ones, leaving larger particles on the surface.

The tendency for falling smaller particles to scatter and leave larger particles on the surface 
has been dubbed 'ballistic sorting' by \citet{shinbrot17}
who  proposed that ballistic sorting of  impactors could account
for large  boulders seen on the surfaces of rubble asteroids. 
Similarly we propose that ballistic sorting
of ejecta launched by an impact generated seismic pulse can strand boulders on an asteroid surface. 
Our experiments show that a single pulse can strand a previously buried large particle.   
We have found that multiple  pulses (but separated in time) continue to unearth larger particles and once a larger particle is on the surface it tends to stay there.

If seismic energy is rapidly attenuated in rubble asteroids then seismic reverberation may be short and ineffective.   
Boulder stranding of ejecta gives an attractive mechanism accounting for large boulders on the surface of asteroids. 
Most mechanisms that produce the Brazil nut effect are effective at low surface gravity but have longer time scales with decreased surface gravity \citep{maurel17,chujo18}.
However, seismic pulses might be effective in low surface gravity environments because they 
can eject more material off the surface.  Future low gravity experiments and simulations might explore this possibility.
While we have primarily considered  sub-catastrophic impacts,
ballistic sorting might happen following catastrophic disruption by an impact and during a phase of re-accumulation.
Stress failure following spin-up 
might also eject equatorial material (e.g., \citealt{sanchez18,yu18})  that could ballistically sort during a phase
of re-accumulation.  

Using  seismic efficiency and source time parameters we explored the regime where  an impact generated
seismic pulse could strand large boulders on the surface of a rubble asteroid such as Bennu or Ryugu.
Our experiments suggested that boulder stranding is  likely to take place if 
 the acceleration in the seismic pulse is
be above that of surface gravity, the ejected material reaches a height above the surface larger than the boulder height, 
and the depth of ejected material is  larger than the boulder height.
For a single impact to leave a large boulder stranded on the surface, we find that the impact must be nearly catastrophic,
the seismic efficiency must be fairly high, $\sim 10^{-4}$, and the seismic pulse duration must be similar
to the sound travel time across the distance of the projectile diameter.
%We assumed that the velocity of an ejected particle depends on the velocity displacement amplitude in the traveling seismic
%pulse and that the depth of ejected material depends on the duration of the pulse.

A single impact excited seismic pulse would be strongest on the surface nearest the impact \citep{thomas05}  and so boulder stranding
by a seismic pulse might give larger boulders near an impact site.  A single impact would be expected to cause 
an inhomogeneous surface size distribution or one that varies as a function of position on the surface.  
If the size particle distribution does not vary with position on the surface (e.g. as suggested by observations
of 4179 Toutatis; \citealt{jiang15}), then one might rule out this process as an explanation for large surface boulders, and we might
place constraints on the nature of seismic pulse propagation in the interior.

The mechanism for boulder stranding
explored here may not explain the correlation of large and small boulders being distributed into regions of high and low potential observed on Itokawa \citep{miyamoto07,tancredi15}. Our process is dependent on the there being large boulders below the surface before a subsurface pulse ejects material. If there are no large boulders below the surface of the low potential regions then only smaller boulders will be present and we are not able to distinguish our process compared to others. However, if there are large boulders present below the surface when material is ejected then our experiments have shown that smaller particles will scatter off large particles during landing. It is possible that when material is ejected off the surface of Itokawa the smaller boulders will scatter into low potential regions leaving larger boulders in the high potential regions. Our mechanism cannot fully explain the correlation of the boulder size distribution with potential on Itokawa suggesting size segregation processes on asteroids likely include more than one mechanism.
%completely lack boulders \citep{miyamoto07},
%so size segregation processes on asteroids likely include more than one mechanism.

\subsection{Role of microgravity and atmosphere}
Our experiments are carried out under the gravitational and atmospheric conditions of Earth (one atmosphere of pressure and 1g) and so do not necessarily mimic conditions or materials on an asteroid.  A milli, or micro, gravity environment would reduce the strength of a pulse  required to eject surface particles and would prolong their time of flight.  However lower gravity would also affect the strength of hydrostatic forces  and so how the pulse propagates through the medium. Any size segregation due to the Brazil Nut effect would still occur, but take a longer time for lower surface gravities.

As particles can be ejected at lower velocities in  reduced-gravity environments, collisions between particles upon landing would also be less energetic. If  the coefficient of restitution is higher than in our experiments, particles would  travel further after scattering facilitating boulder stranding.

As evidenced by the sample from asteroid Itokawa, asteroid regolith is likely to contain very fine particles (dust). In our experiments however, we have deliberately removed these particles in order to mitigate the  role of attenuation caused by air drag.  Additionally, the presence of dust would necessarily introduce cohesive forces into the dynamics of the system, something we have neglected so that we can identify and understand the main mechanism behind our experimental results and their application to asteroids.

We have also neglected attenuation of a seismic pulse when estimating a  regime for boulder stranding on an asteroid, but let the energy spread out as  the inverse distance squared. However, attenuation and scattering probably  would broaden the pulse and reduce its amplitude. The shape of  the pulse through time, the wave speed as a function of depth in the asteroid, the   pulse arrival time, and strength at different locations on the surface, would also influence the ejecta velocity distribution and should be investigated.  However, we have left them out of this first set of experiments and analysis in order to maintain their simplicity.  Future research will attempt to include all these other variables and study their influence.

\vskip 2 truein
------------------------------
Acknowledgements

We would like to thank L'Observatoire de la C\^ote d'Azur for their warm welcome,
and hospitality March and April 2018.
We thank Hesam Askari, Yuhei Zhao and Ferdinand Weisenhorn for helpful discussions,
and Gordon Rice for providing tips and materials for lighting our videos.

We are grateful for generous support from the Simons Foundation 2017--2018.
This material is based upon work supported in part supported by NASA grant 80NSSC17K0771,
National Science Foundation Grant No. PHY-1757062 and  
NASA grant NNX15AI46G (PGGURP) to PI Tracy Gregg. 

S.R.S. acknowledges support from NASA Grant no. 80NSSC18K0226 as part of the OSIRIS-REx Participating Scientist Program as well as the Complex Systems and Space, Environment, Risk and Resilience Academies of the Initiative d'EXcellence "Joint, Excellent, and Dynamic Initiative" (IDEX JEDI) of the Universit\/e C\^ote d'Azur.

\vskip 2 truein
------------------------------
References
\bibliographystyle{elsarticle-harv}
\bibliography{refs_boulders}

\begin{thebibliography}{57}
\expandafter\ifx\csname natexlab\endcsname\relax\def\natexlab#1{#1}\fi
\expandafter\ifx\csname url\endcsname\relax
  \def\url#1{\texttt{#1}}\fi
\expandafter\ifx\csname urlprefix\endcsname\relax\def\urlprefix{URL }\fi

\bibitem[{Allan et~al.(2016)Allan, Caswell, Keim, and van~der Wel}]{trackpy}
Allan, D., Caswell, T., Keim, N., van~der Wel, C., August 2016. Trackpy v0.3.2.
\newline\urlprefix\url{https://doi.org/10.5281/zenodo.60550}

\bibitem[{Asphaug(2008)}]{asphaug08}
Asphaug, E., 2008. Critical crater diameter and asteroid impact seismology.
  Meteoritics \& Planetary Science 43~(6), 1075--1084.

\bibitem[{Asphaug et~al.(2002)Asphaug, Ryan, and T.~Zuber}]{asphaug02}
Asphaug, E., Ryan, E., T.~Zuber, M., 01 2002. Asteroid Interiors.

\bibitem[{Bottke et~al.(1994)Bottke, Nolan, Greenberg, and Kolvoord}]{bottke94}
Bottke, W.~F., Nolan, M., Greenberg, R., Kolvoord, R., 1994. Velocity
  distributions among colliding asteroids. Icarus 107, 255--268.

\bibitem[{Cheng et~al.(2002)Cheng, Izenberg, Chapman, and Zuber}]{cheng02}
Cheng, A., Izenberg, N., Chapman, C., Zuber, M.~T., 2002. Ponded deposits on
  asteroid 433 eros. Meteoritics \& Planetary Science 37, 1095--1105.

\bibitem[{Chujo et~al.(2018)Chujo, Mori, Kawaguchi, and Yano}]{chujo18}
Chujo, T., Mori, O., Kawaguchi, J., Yano, H., 2018. Categorization of brazil
  nut effect and its reverse under less-convective conditions for microgravity
  geology. Monthly Notices of the Royal Astronomical Society, 474~(4),
  4447--4459.

\bibitem[{Cintala et~al.(1978)Cintala, Head, and Veverka}]{cintala78}
Cintala, M.~J., Head, J.~W., Veverka, J., 1978. Characteristics of the
  cratering process on small satellites and asteroids. Proceedings of Lunar
  Sci. Conf. 9, 3803--3830.

\bibitem[{Clement et~al.(2010)Clement, Pacheco-Martinez, Swift, and
  King}]{clement10}
Clement, C.~P., Pacheco-Martinez, H.~A., Swift, M.~R., King, P.~J., sep 2010.
  The water-enhanced brazil nut effect. {EPL} (Europhysics Letters) 91~(5),
  54001.
\newline\urlprefix\url{https://doi.org/10.1209%2F0295-5075%2F91%2F54001}

\bibitem[{Crocker and Grier(1996)}]{crocker96}
Crocker, J.~C., Grier, D.~G., 1996. Methods of digital video microscopy for
  colloidal studies. Journal of Colloid Interface Science 179, 298--310.

\bibitem[{Dainty et~al.(1974)Dainty, Toks\"oz, Anderson, Pines, Nakamura, and
  Latham}]{dainty74}
Dainty, A., Toks\"oz, M.~N., Anderson, K., Pines, P., Nakamura, Y., Latham, G.,
  1974. Seismic scattering and shallow structure of the moon in oceanus
  procellarum. Moon 9, 11--29.

\bibitem[{Greenberg et~al.(1996)Greenberg, Bottke, M., Geissler, Petit, Durda,
  Asphaug, and Head}]{greenberg96}
Greenberg, R., Bottke, W., M., N., Geissler, P., Petit, J., Durda, D., Asphaug,
  E., Head, J., 1996. Collisional and dynamical history of ida. Icarus 120,
  106--118.

\bibitem[{Greenberg et~al.(1994)Greenberg, Nolan, Bottke, Kolvoord, and
  Veverka}]{greenberg94}
Greenberg, R., Nolan, M., Bottke, W., Kolvoord, R., Veverka, J., 1994.
  Collisional history of gaspra. Icarus 107, 84--97.

\bibitem[{G\"uldemeister and W\"unnemann(2017)}]{guldemeister17}
G\"uldemeister, N., W\"unnemann, K., 2017. Quantitative analysis of
  impact-induced seismic signals by numerical modeling. Icarus 296, 15--27.

\bibitem[{Herrmann et~al.(1997)Herrmann, Hovi, and Luding}]{jenkins97}
Herrmann, H.~J., Hovi, J.~P., Luding, S. (Eds.), 1997. Physics of Dry Granular
  Media. Vol. Vol. 350 of NATO ASI Series. Kluwer Academic Publishers,
  Dordrecht.

\bibitem[{Hirata and Nakamura(2006)}]{hirata06}
Hirata, N., Nakamura, A.~M., 2006. Secondary craters of tycho: Size-frequency
  distributions and estimated fragment size-velocity relationships. Journal of
  Geophysical Research 111~(E3), E03005.

\bibitem[{Hoerth et~al.(2014)Hoerth, Schafer, Nau, Kuder, Poelchau, Thoma, and
  Kenkmann}]{hoerth14}
Hoerth, T., Schafer, F., Nau, S., Kuder, J., Poelchau, M.~H., Thoma, K.,
  Kenkmann, T., 2014. In situ measurements of impact-induced pressure waves in
  sandstone targets. Journal of Geophysical Research Planets 119, 2177--2187.

\bibitem[{Holsapple(1993)}]{holsapple93}
Holsapple, K.~A., 1993. The scaling of impact processes in planetary sciences.
  Annual Review of Earth and Planetary Sciences 21, 333--373.

\bibitem[{Hong et~al.(2001)Hong, Quinn, and Luding}]{hong01}
Hong, D.~C., Quinn, P.~V., Luding, S., 2001. Reverse brazil nut problem:
  Competition between percolation and condensation. Physics Review Letters
  86~(15), 3423--3426.

\bibitem[{Housen and Holsapple(2011)}]{housen11}
Housen, K.~R., Holsapple, K.~A., 2011. Ejecta from impact craters. Icarus 211,
  856--875.

\bibitem[{Jiang et~al.(2015)Jiang, Ji, Huang, Marchi, Li, and Ip}]{jiang15}
Jiang, Y., Ji, J., Huang, J., Marchi, S., Li, Y., Ip, W.-H., 11 2015. Boulders
  on asteroid toutatis as observed by chang'e-2. Scientific Reports 5, 16029
  EP.

\bibitem[{Jullien et~al.(1992)Jullien, Meakin, and Pavlovitch}]{jullien92}
Jullien, R., Meakin, P., Pavlovitch, A., 1992. Three-dimensional model for
  particle-size segregation by shaking. Physical Review Letters 69~(4),
  .640--643.

\bibitem[{Jutzi et~al.(2010)Jutzi, Michel, Benz, and Richardson}]{jutzi10}
Jutzi, M., Michel, P., Benz, W., Richardson, D.~C., 2010. Fragment properties
  at the catastrophic disruption threshold: The effect of the parent body's
  internal structure. Icarus 207, 54--65.

\bibitem[{Knight et~al.(1993)Knight, Jaeger, and Nagel}]{knight93}
Knight, J., Jaeger, H., Nagel, S., 1993. Vibration-induced size separation in
  granular media: The convection connection. Physics Review Letters 70,
  3728--3731.

\bibitem[{Latham et~al.(1970)Latham, McDonald, and Moore}]{latham70}
Latham, G.~V., McDonald, W.~G., Moore, H.~J., 1970. Missile impacts as sources
  of seismic energy on the moon. Science 168~(3928), 242--245.

\bibitem[{Lognonn\'e et~al.(2009)Lognonn\'e, Feuvre, Johnson, and
  Weber}]{lognonne09}
Lognonn\'e, P., Feuvre, M.~L., Johnson, C.~L., Weber, R.~C., 2009. Moon
  meteoritic seismic hum: steady state prediction. Journal Geophysical Research
  (Planets) 114, 12003.

\bibitem[{Makse et~al.(2004)Makse, Gland, Johnson, and Schwartz}]{makse04}
Makse, H.~A., Gland, N., Johnson, D.~L., Schwartz, L., 2004. Granular packings:
  Nonlinear elasticity, sound propagation, and collective relaxation dynamics.
  Physical Review E 70, 061302.

\bibitem[{Matsumura et~al.(2014)Matsumura, Richardson, Michel, Schwartz, and
  Ballouz}]{matsumura14}
Matsumura, S., Richardson, D.~C., Michel, P., Schwartz, S.~R., Ballouz, R.-L.,
  2014. The brazil nut effect and its application to asteroids. Monthly Notices
  of the Royal Astronomical Society 443, 3368--3380.

\bibitem[{Maurel et~al.(2017)Maurel, Ballouz, Richardson, Michel, and
  Schwartz}]{maurel17}
Maurel, C., Ballouz, R.-L., Richardson, D.~C., Michel, P., Schwartz, S.~R.,
  2017. Numerical simulations of oscillation-driven regolith motion: Brazil-nut
  effect. Monthly Notices of the Royal Astronomical Society 464, 2866--2881.

\bibitem[{McGarr et~al.(1969)McGarr, Latham, and Gault}]{mcgarr69}
McGarr, A., Latham, G., Gault, D., 1969. Meteoroid impacts as sources of
  seismicity on the moon. Journal of Geophysical Research 74, 5981--5994.

\bibitem[{Meakin et~al.(1986)Meakin, Ramanlal, Sander, and Ball}]{meakin86}
Meakin, P., Ramanlal, P., Sander, L.~M., Ball, R.~C., 1986. Ballistic
  deposition on surfaces. Physical Review A 34, 5091.

\bibitem[{Melosh(1989)}]{melosh89}
Melosh, H.~J., 1989. Impact cratering: a geologic process. Oxford Monographs on
  Geology and Geophysics. Oxford University Press, Oxford England.

\bibitem[{Miyamoto et~al.(2007)Miyamoto, Yano, Scheeres, Abe, Barnouin-Jha,
  Cheng, Demura, Gaskell, Hirata, Ishiguro, Michikami, Nakamura, Nakamura,
  Saito, and Sasaki}]{miyamoto07}
Miyamoto, H., Yano, H., Scheeres, D.~J., Abe, S., Barnouin-Jha, O., Cheng,
  A.~F., Demura, H., Gaskell, R.~W., Hirata, N., Ishiguro, M., Michikami, T.,
  Nakamura, A.~M., Nakamura, R., Saito, J., Sasaki, S., May 2007. Regolith
  migration and sorting on asteroid itokawa. Science 316~(5827), 1011--1014.

\bibitem[{M\"obius et~al.(2005)M\"obius, Cheng, Eshuis, Karczmar, Nagel, and
  Jaeger}]{mobius05}
M\"obius, M.~E., Cheng, X., Eshuis, P., Karczmar, G.~S., Nagel, S.~R., Jaeger,
  H.~M., Jul 2005. Effect of air on granular size separation in a vibrated
  granular bed. Phys. Rev. E 72, 011304.
\newline\urlprefix\url{https://link.aps.org/doi/10.1103/PhysRevE.72.011304}

\bibitem[{Nakamura(1976)}]{nakamura76}
Nakamura, Y., 1976. Seismic energy transmission in the lunar surface zone
  determined from signals generated by movement of lunar rovers. Bulletin of
  the Seismological Society of America 66~(2), 593--606.

\bibitem[{Naylor et~al.(2003)Naylor, Swift, and King}]{naylor03}
Naylor, M.~A., Swift, M.~R., King, P.~J., Jul 2003. Air-driven brazil nut
  effect. Phys. Rev. E 68, 012301.
\newline\urlprefix\url{https://link.aps.org/doi/10.1103/PhysRevE.68.012301}

\bibitem[{{Nolan} et~al.(1992){Nolan}, {Asphaug}, and {Greenberg}}]{nolan92}
{Nolan}, M.~C., {Asphaug}, E., {Greenberg}, R., Jun. 1992. {Numerical
  Simulation of Impacts on Small Asteroids}. In: AAS/Division for Planetary
  Sciences Meeting Abstracts \#24. Vol.~24 of Bulletin of the American
  Astronomical Society. p. 959.

\bibitem[{O'Donovan et~al.(2016)O'Donovan, Ibraim, O'Sullivan, Hamlin, Wood,
  and Marketos}]{odonovan16}
O'Donovan, J., Ibraim, E., O'Sullivan, C., Hamlin, S., Wood, D.~M., Marketos,
  G., 2016. Micromechanics of seismic wave propagation in granular materials.
  Granular Matter 18, 56.

\bibitem[{Perera et~al.(2016)Perera, Jackson, Asphaug, and Ballouz}]{perera16}
Perera, V., Jackson, A.~P., Asphaug, E., Ballouz, R.-L., 2016. The spherical
  brazil nut effect and its significance to asteroids. Icarus 278, 194--203.

\bibitem[{Richardson et~al.(2004)Richardson, Melosh, and
  Greenberg}]{richardson04}
Richardson, J.~E., Melosh, H.~J., Greenberg, R., 2004. Impact-induced seismic
  activity on asteroid 433eros: A surface modification process. Science 306,
  1526--1529.

\bibitem[{Richardson et~al.(2005)Richardson, Melosh, Greenberg, and
  O'Brien}]{richardson05}
Richardson, J.~J., Melosh, H.~J., Greenberg, R.~J., O'Brien, D.~P., 2005. The
  global effects of impact-induced seismic activity on fractured asteroid
  surface morphology. Icarus 179, 325--349.

\bibitem[{Rosato et~al.(1987)Rosato, Strandburg, Prinz, and
  Swendsen}]{rosato87}
Rosato, A., Strandburg, K.~J., Prinz, F., Swendsen, R.~H., 1987. Why the brazil
  nuts are on top: Size segregation of particulate matter by shaking. Physics
  Review Letters 58, 1038--1042.

\bibitem[{S\'anchez and Scheeres(2014)}]{sanchez14}
S\'anchez, P., Scheeres, D.~J., May 2014. The strength of regolith and rubble
  pile asteroids. Meteoritics \& Planetary Science 49~(5), 788--811.

\bibitem[{Sanchez and Scheeres(2018)}]{sanchez18}
Sanchez, P., Scheeres, D.~J., 2018. Rotational evolution of self-gravitating
  aggregates with cores of variable strength. Planetary and Space Science 157,
  39--47.

\bibitem[{Scheeres and Sanchez(2018)}]{scheeres18}
Scheeres, D.~J., Sanchez, P., 2018. Implications of cohesive strength in
  asteroid interiors and surfaces and its measurement. Progress in Earth and
  Planetary Science 5, 25.

\bibitem[{Schultz and Gault(1975)}]{schultz75}
Schultz, P.~H., Gault, D.~E., 1975. Seismic effects from major basin formations
  on the moon and mercury. Moon 12, 159--177.

\bibitem[{Shinbrot et~al.(2017)Shinbrot, Sabuwala, Siu, Lazo, and
  Chakraborty}]{shinbrot17}
Shinbrot, T., Sabuwala, T., Siu, T., Lazo, M.~V., Chakraborty, P., 2017. Size
  sorting on the rubble-pile asteroid, itokawa,. Physics Reviews Letters 118,
  111101.

\bibitem[{Shishkin(2007)}]{shishkin07}
Shishkin, N.~I., 2007. Seismic efficiency of a contact explosion and a high
  velocity impact. Journal of Applied Mechanics and Technical Physics 48,
  145--152.

\bibitem[{Tancredi et~al.(2012)Tancredi, Maciel, Heredia, Richeri, and
  Nesmachnow}]{tancredi12}
Tancredi, G., Maciel, A., Heredia, L., Richeri, P., Nesmachnow, S., 2012.
  Granular physics in low-gravity environments using discrete element method.
  Monthly Notices of the Royal Astronomical Society 420, 3368--3380.

\bibitem[{Tancredi et~al.(2015)Tancredi, Roland, and Bruzzone}]{tancredi15}
Tancredi, G., Roland, S., Bruzzone, S., 2015. Distribution of boulders and the
  gravity potential on asteroid itokawa. Icarus 247, 279--290.

\bibitem[{Thomas and Robinson(2005)}]{thomas05}
Thomas, P.~C., Robinson, M.~S., 2005. Seismic resurfacing by a single impact on
  the asteroid 433 eros. Nature 436, 366--369.

\bibitem[{Titley(1966)}]{titley66}
Titley, S.~R., 1966. Seismic energy as an agent of morphologic modification on
  the moon. Tech. rep., United States Geological Survey, Flagstaff, AZ.

\bibitem[{Toks\"oz et~al.(1974)Toks\"oz, Dainty, Solomon, and
  Anderson}]{toksoz74}
Toks\"oz, M.~N., Dainty, A.~M., Solomon, S., Anderson, K., 1974. Structure of
  the moon. Reviews of Geophysics 12~(4), 539--567.

\bibitem[{Williams(1976)}]{williams76}
Williams, J., 1976. The segregation of particulate materials. a review. Powder
  Technology 15~(2), 245--251.

\bibitem[{Wolf(1944)}]{wolf44}
Wolf, A., 1944. The equation of motion of a geophone on the surface of an
  elastic earth. Geophysics 9, 29--34.

\bibitem[{Yamada et~al.(2016)Yamada, Ando, Morota, and Katsuragi}]{yamada16}
Yamada, T.~M., Ando, K., Morota, T., Katsuragi, H., 2016. Timescale of asteroid
  resurfacing by regolith convection resulting from the impact-induced global
  seismic shaking. Icarus 272, 165--177.

\bibitem[{Yasui et~al.(2015)Yasui, Matsumoto, and Arakawa}]{yasui15}
Yasui, M., Matsumoto, E., Arakawa, M., 2015. Experimental study on
  impact-induced seismic wave propagation through granular materials. Icarus
  260, 320--331.

\bibitem[{Yu et~al.(2018)Yu, Michel, Hirabayashi, Schwartz, Zhang, Richardson,
  and Liu}]{yu18}
Yu, Y., Michel, P., Hirabayashi, M., Schwartz, S.~R., Zhang, Y., Richardson,
  D.~C., Liu, X., 2018. The dynamical complexity of surface mass shedding from
  a top-shaped asteroid near the critical spin limit. Astronomical Journal 156,
  59--77.

\end{thebibliography}

\end{document}